\documentclass[aps,prl,showpacs,superscriptaddress,groupedaddress,nofootinbib]{revtex4-2}  % for review and submission
\usepackage{cancel}
\usepackage{graphicx} % Required for inserting images
\usepackage[T1]{fontenc}
\usepackage{lmodern} 
\usepackage{graphicx}  % needed for figures
\usepackage{dcolumn}   % needed for some tables
\usepackage{bm}        % for math
\usepackage{amssymb}   % for math
\usepackage{amsmath}
\usepackage{bbold}
\usepackage{siunitx}
\usepackage{array}
\usepackage{float}
\usepackage{makecell}
\usepackage{braket}
\usepackage{longtable}
\usepackage{supertabular,booktabs}
\usepackage{soul}

\usepackage{titlesec} %for hyperref to detect names of sections etc
\usepackage[colorlinks,citecolor=blue,urlcolor=blue,hypertexnames=true]{hyperref}
\usepackage{subfigure}
\usepackage[usenames,dvipsnames,svgnames,table]{xcolor} % use color 

\renewcommand{\arraystretch}{2}

\newcommand{\be}{\begin{equation}}
\newcommand{\ee}{\end{equation}}
\newcommand{\bea}{\begin{eqnarray}}
\newcommand{\eea}{\end{eqnarray}}
\newcommand{\bml}{\begin{subequations}}
\newcommand{\eml}{\end{subequations}}
\newcommand{\bfig}{\begin{figure}}
\newcommand{\efig}{\end{figure}}

\newcommand{\slashed}{\hspace{-1.1ex}/}

\newcommand{\bmat}{\begin{pmatrix}}
\newcommand{\emat}{\end{pmatrix}}
\usepackage{graphicx, slashed,booktabs, color, multirow, float,
amsfonts, bbold, mathtools, sidecap, tikz, bm,enumitem}
\usepackage{multirow}
\usepackage{bbding}
\usepackage{titlesec}
\usepackage{hyperref}
\usepackage{wasysym}
\usepackage{amssymb}% http://ctan.org/pkg/amssymb
\usepackage{pifont}% http://ctan.org/pkg/pifont

\renewcommand{\geq}{\geqslant}

\usepackage[dvipsnames, usenames]{xcolor}

\definecolor{linkcolor}{rgb}{0.55, 0.13, .32}

\definecolor{oucrimsonred}{rgb}{0.6, 0.0, 0.0}
\definecolor{persianblue}{rgb}{0.11, 0.22, 0.73}
\definecolor{forestgreen}{rgb}{0.13,0.35,0.13}
\definecolor{lightgray}{rgb}{0.83, 0.83, 0.83}
 \hypersetup{colorlinks, citecolor=oucrimsonred, linkcolor=persianblue, urlcolor=oucrimsonred}
\definecolor{cornellred}{rgb}{0.7, 0.11, 0.11}
\definecolor{navyblue}{rgb}{0.0, 0.0, 0.5}
\definecolor{amethyst}{rgb}{0.6, 0.4, 0.8}
\definecolor{yellow}{rgb}{1.0, 1.0, 0.0}
\definecolor{firebrick}{rgb}{0.7, 0.13, 0.13}
\definecolor{tangerineyellow}{rgb}{1.0, 0.8, 0.0}
\definecolor{deepfuchsia}{rgb}{0.76, 0.33, 0.76}
\definecolor{amber}{rgb}{1.0, 0.75, 0.0}
\definecolor{VioletRed4}{rgb}{0.55, 0.13, .32}
\definecolor{indiagreen}{rgb}{0.07, 0.53, 0.03}
\definecolor{VioletRed4}{rgb}{0.55, 0.13, .32}

\usepackage{hyperref}
\usepackage{graphics, appendix, afterpage, makecell} 
\usepackage{bbold}
\usepackage{tikz}
\usepackage{adjustbox}

\usepackage{tcolorbox}

\def\Re{\mbox{Re}\,}

\definecolor{oucrimsonred}{rgb}{0.6, 0.0, 0.0}
\definecolor{persianblue}{rgb}{0.11, 0.22, 0.73}
\definecolor{forestgreen}{rgb}{0.13,0.35,0.13}
\definecolor{lightgray}{rgb}{0.83, 0.83, 0.83}
 \hypersetup{colorlinks, citecolor=oucrimsonred, linkcolor=persianblue, urlcolor=oucrimsonred}
\definecolor{cornellred}{rgb}{0.7, 0.11, 0.11}
\definecolor{navyblue}{rgb}{0.0, 0.0, 0.5}
\definecolor{amethyst}{rgb}{0.6, 0.4, 0.8}
\definecolor{yellow}{rgb}{1.0, 1.0, 0.0}
\definecolor{firebrick}{rgb}{0.7, 0.13, 0.13}
\definecolor{tangerineyellow}{rgb}{1.0, 0.8, 0.0}
\definecolor{deepfuchsia}{rgb}{0.76, 0.33, 0.76}
\definecolor{amber}{rgb}{1.0, 0.75, 0.0}
\definecolor{VioletRed4}{rgb}{0.55, 0.13, .32}
\definecolor{indiagreen}{rgb}{0.07, 0.53, 0.03}
\definecolor{VioletRed4}{rgb}{0.55, 0.13, .32}

\definecolor{oucrimsonred}{rgb}{0.6, 0.0, 0.0}
\newcommand\vertarrowbox[3][6ex]{%
  \begin{array}[t]{@{}c@{}} #2 \\
  \left\uparrow\vcenter{\hrule height #1}\right.\kern-\nulldelimiterspace\\
  \makebox[0pt]{\scriptsize#3}
  \end{array}%
}

\definecolor{mtcolor}{rgb}{.8,.3,.1}

\definecolor{violachiaro}{rgb}{1,0.6,1}

\definecolor{gbcolor}{rgb}{.43,.22,.12}
 
\definecolor{gbcolor2}{rgb}{.9,.2,.6}
\definecolor{gbcolor3}{rgb}{.3,.2,.6}

\definecolor{verdechiaro}{rgb}{0.6,1,0.6}
\definecolor{giallochiaro}{rgb}{1,1,0.6}
\definecolor{bluscuro}{rgb}{0.15, 0.2, 0.9}
\definecolor{verdes}{rgb}{0.1, 0.5, 0.1}%
\definecolor{tangerineyellow}{rgb}{1.0, 0.8, 0.0}
\definecolor{smokyblack}{rgb}{0.06, 0.05, 0.03}

\definecolor{americanrose}{rgb}{1.0, 0.01, 0.24}
\definecolor{cobalt}{rgb}{0.0, 0.28, 0.67}
\definecolor{brandeisblue}{rgb}{0.0, 0.44, 1.0}
\definecolor{mycolor}{rgb}{0.0, 0.0, 0.5}%navyblue
\definecolor{oxfordblue}{rgb}{0.0, 0.13, 0.28}
\definecolor{azure}{rgb}{0.0, 0.5, 1.0}
\definecolor{turquoiseblue}{rgb}{0.0, 1.0, 0.94}
\newtcolorbox{mynewbox}[1]{colback=white!5!white,colframe=azure!75!black,fonttitle=\bfseries,title=#1}
\newtcolorbox{mybox}{colback=mycolor!5!white,colframe=azure!75!black}
\newtcolorbox{mynamedbox}[1]{colback=mycolor!5!white,colframe=azure!75!black,title=#1}
\definecolor{venetianred}{rgb}{0.78, 0.03, 0.08}
\newtcolorbox{mynamedbox1}[1]{colback=venetianred!5!white,colframe=venetianred!80!black,title=#1}
\newtcolorbox{mynamedbox2}[1]{colback=azure!5!white,colframe=azure!80!black,title=#1}

\definecolor{rossocorsa}{rgb}{0.83, 0.0, 0.0}

\tikzset{->-/.style={decoration={
  markings,
  mark=at position #1 with {\arrow{>}}},postaction={decorate}}}
\tikzset{-<-/.style={decoration={
  markings,
  mark=at position #1 with {\arrow{<}}},postaction={decorate}}} 
\def\be{\begin{equation}}
\def\ee{\end{equation}}
\def\ba{\begin{eqnarray}}
\def\ea{\end{eqnarray}}

\def\L*{{\cal L}_*}
\def\L{\mathcal{L}}

\def\<{\langle}
\def\>{\rangle}

 \def\neq {\not\equiv}

\def\cs2{c_{s}^{2}}

 \def\be   {\begin{equation}}   \def\ee   {\end{equation}}
 \def\ba   {\begin{array}}      \def\ea   {\end{array}}
 \def\bea  {\begin{eqnarray}}   \def\eea  {\end{eqnarray}}
 \def\bean {\begin{eqnarray*}}  \def\eean {\end{eqnarray*}}
\titleclass{\subsubsubsection}{straight}[\subsection]

\newcounter{subsubsubsection}[subsubsection]
\renewcommand\thesubsubsubsection{\thesubsubsection.\arabic{subsubsubsection}}
\titleformat{\subsubsubsection}
  {\normalfont\normalsize\bfseries}{\thesubsubsubsection}{1em}{}
\titlespacing*{\subsubsubsection}
{0pt}{3.25ex plus 1ex minus .2ex}{1.5ex plus .2ex}

\makeatletter
\renewcommand\paragraph{\@startsection{paragraph}{5}{\z@}%
  {3.25ex \@plus1ex \@minus.2ex}%
  {-1em}%
  {\normalfont\normalsize\bfseries}}
\renewcommand\subparagraph{\@startsection{subparagraph}{6}{\parindent}%
  {3.25ex \@plus1ex \@minus .2ex}%
  {-1em}%
  {\normalfont\normalsize\bfseries}}
\def\toclevel@subsubsubsection{4}
\def\toclevel@paragraph{5}
\def\toclevel@paragraph{6}
\def\l@subsubsubsection{\@dottedtocline{4}{7em}{4em}}
\def\l@paragraph{\@dottedtocline{5}{10em}{5em}}
\def\l@subparagraph{\@dottedtocline{6}{14em}{6em}}
\makeatother

\usepackage{titlesec} %for hyperref to detect names of sections etc
\usepackage[colorlinks,citecolor=blue,urlcolor=blue,hypertexnames=true]{hyperref}
\usepackage{subfigure}
\usepackage[usenames,dvipsnames,svgnames,table]{xcolor}

\usepackage{tikzsymbols}
\usepackage{natbib}
\usepackage{float}

\usepackage{tikz,xcolor,hyperref}

\usepackage{slashed}

\definecolor{lime}{HTML}{A6CE39}
\DeclareRobustCommand{\orcidicon}{
	\begin{tikzpicture}
	\draw[lime, fill=lime] (0,0) 
	circle [radius=0.2] 
	node[white] {{\fontfamily{qag}\selectfont \tiny ID}};
	\draw[white, fill=white] (-0.0625,0.095) 
	circle [radius=0.007];
	\end{tikzpicture}
	\hspace{-2mm}
}

\usepackage[usenames,dvipsnames,svgnames,table]{xcolor}  
\usepackage{tikzsymbols}
\usepackage{natbib}
\usepackage{float}

\usepackage{tikz,xcolor,hyperref}

\usepackage{slashed}

\definecolor{lime}{HTML}{A6CE39}
\DeclareRobustCommand{\orcidicon}{
	\begin{tikzpicture}
	\draw[lime, fill=lime] (0,0) 
	circle [radius=0.2] 
	node[white] {{\fontfamily{qag}\selectfont \tiny ID}};
	\draw[white, fill=white] (-0.0625,0.095) 
	circle [radius=0.007];
	\end{tikzpicture}
	\hspace{-2mm}
}

\foreach \x in {A, ..., Z}{\expandafter\xdef\csname orcid\x\endcsname{\noexpand\href{https://orcid.org/\csname orcidauthor\x\endcsname}
			{\noexpand\orcidicon}}
}
\usepackage{bbold}
\usepackage{tikz}
\usepackage{adjustbox}
\usepackage{tcolorbox}
\usepackage{enumitem}
\usepackage{amsfonts}

\setlist[itemize,1]{label=$\times$}
\setlist[itemize,2]{label=$\checkmark$}
\setlist[itemize,3]{label=$\diamond$}
\setlist[itemize,4]{label=$\bullet$}

\begin{document}
\title{\Large \textcolor{Sepia}{Three-Loop Gauge Beta Functions in Supersymmetric Theories with Exponential Higher Covariant Derivative Regularization}}
\author{\large Swapnil Kumar Singh\orcidF{}${}^{1}$}
\email{swapnilsingh.ph@gmail.com (Corresponding Author)}
\affiliation{ ${}^{1}$B.M.S. College of Engineering, 
    Bangalore, Karnataka, 560019, India.}

\begin{abstract}
We study the three-loop gauge $\beta$-functions in general $\mathcal{N}=1$ supersymmetric gauge theories regularized by higher covariant derivatives (HCD) supplemented with Pauli--Villars subtraction. The all-structure three-loop form of the $\beta$-functions in the HCD framework is known and involves regulator-dependent parameters. Here we evaluate these parameters explicitly for the exponential regulators $R(x)=e^{x^n}$ and $F(x)=e^{x^m}$. We obtain the constants $A(n)$ and $B(m)$ in closed form, together with their large-$n,m$ asymptotics, and substitute them into the general three-loop expressions. This yields fully explicit, regulator-parameterized $\beta$-functions and a systematic expansion in $1/n$ and $1/m$ that organizes finite, scheme-dependent terms. We then exhibit finite coupling redefinitions that map the renormalized $\overline{\mathrm{DR}}$ result to a scheme compatible with the Novikov--Shifman--Vainshtein--Zakharov relation. Our analysis clarifies how exponential higher-derivative regulators preserve this relation at the bare level and illustrates the regulator-driven structure of supersymmetric renormalization group flows.

\end{abstract}

\maketitle

% --- Introduction ---
\section{Introduction}
\label{sec:Introduction}

The study of renormalization group (RG) functions in $\mathcal{N}=1$ supersymmetric gauge theories is central to understanding both perturbative and non-perturbative aspects of quantum field theory. In particular, the gauge $\beta$-functions, which determine the scale dependence of gauge couplings, serve as a bridge between high-energy unification, low-energy phenomenology, and the internal consistency of supersymmetric effective field theories~\cite{Ellis1991,Amaldi1991,Langacker1991}. Precision knowledge of these functions is indispensable for testing supersymmetric extensions of the Standard Model, constraining scenarios of Grand Unified Theories (GUTs), and analyzing dualities in strongly coupled regimes.

At the one- and two-loop levels the gauge $\beta$-functions are well established~\cite{Jones1983,Novikov1986}. These results form the foundation for phenomenological applications, including the classic demonstration that gauge couplings unify in the Minimal Supersymmetric Standard Model (MSSM) near $10^{16}\,\text{GeV}$~\cite{Ellis1991,Amaldi1991,Langacker1991}. While three-loop corrections are numerically smaller, they are essential for achieving percent-level precision in unification fits, refining proton decay predictions, and improving constraints on the superpartner spectrum~\cite{Jack2005,Kazakov1999}. It is important to emphasize that higher-loop effects do not shift the unification scale down to the $\mathcal{O}(1$--$10)\,\text{TeV}$ range, but instead correct the matching conditions at the conventional GUT scale. Thresholds at the TeV scale are associated with superpartner masses, and their effects must be carefully separated from genuine high-scale contributions in phenomenological analyses.

A remarkable feature of supersymmetric gauge theories is the Novikov--Shifman--Vainshtein--Zakharov (NSVZ) relation~\cite{Novikov1983,Shifman1986}, which provides an exact all-order formula for the gauge $\beta$-function in terms of group invariants and anomalous dimensions of matter fields. Originally derived using holomorphy, instanton calculus, and anomaly arguments~\cite{Shifman1996,Seiberg1994,ArkaniHamed1997}, the NSVZ relation was later confirmed diagrammatically in supersymmetry-preserving schemes~\cite{Stepanyantz2020}. Its canonical form reads
\begin{equation}
\label{eq:NSVZ_intro}
\begin{split}
  \frac{\beta_K(\alpha,\lambda)}{\alpha_K^2}
  &= -\,\frac{1}{2\pi\left(1 - \tfrac{C_2(G_K)\,\alpha_K}{2\pi}\right)} \\
  &\quad \times \left[3\,C_2(G_K)
     - \sum_a T_{aK}\left(1 - \gamma_a{}^a(\alpha,\lambda)\right)\right],
\end{split}
\end{equation}
where $C_2(G_K)$ is the quadratic Casimir of the gauge group $G_K$, $T_{aK}$ denotes the Dynkin index of the chiral multiplet $\Phi_a$, and $\gamma_a{}^a$ is its anomalous dimension. Equation~\eqref{eq:NSVZ_intro} encapsulates the holomorphic structure of supersymmetric RG flows and highlights the deep connection between supersymmetry, anomalies, and renormalization.

Whether the NSVZ relation is preserved in explicit multi-loop calculations depends crucially on the choice of regularization and subtraction scheme. Dimensional reduction ($\overline{\text{DR}}$), although widely used in phenomenology, does not preserve the NSVZ form beyond two loops without finite redefinitions of couplings~\cite{Jack1996_1,Jack1996_2}. By contrast, the higher covariant derivative (HCD) regularization introduced by Slavnov~\cite{Slavnov1971,Slavnov1972} has proven to be especially powerful. In this framework, higher-derivative operators suppress ultraviolet divergences, while residual one-loop divergences are canceled using Pauli--Villars (PV) superfields~\cite{Slavnov1977,Kataev2013}. This method preserves both gauge invariance and supersymmetry, and when RG functions are defined in terms of bare couplings, the NSVZ relation holds exactly~\cite{Stepanyantz2020}.

Recent advances have pushed these results to the three-loop level for general 
$\mathcal{N}=1$ supersymmetric theories with semi-simple gauge groups, Yukawa couplings, 
and HCD regularization~\cite{Kazantsev2020,Haneychuk2022,Haneychuk2025}. 
These computations demonstrate explicit consistency with the NSVZ relation and introduce 
two regulator-dependent constants, $A$ and $B$, defined in 
Eq.~\eqref{eq:A_B_constants_multiline}. 
Here $R(x)$ and $F(x)$ denote the regulator functions in the gauge and matter sectors, 
respectively. These constants encode finite, scheme-dependent contributions to higher-loop 
RG functions. Their explicit evaluation is therefore indispensable for connecting formal 
multi-loop results with physical predictions.

The focus of the present paper is to compute $A$ and $B$ explicitly for the family of exponential regulators
\begin{equation}
\label{eq:RegChoice_intro}
R(x) = e^{x^n}, 
\qquad 
F(x) = e^{x^m}, 
\qquad n,m \geq 2,
\end{equation}
which provide strong ultraviolet suppression and analytic control. We demonstrate that
\begin{equation}
\label{eq:AB_result_intro}
A(n) = \frac{\gamma_E}{n}, 
\qquad 
B(m) = \frac{\gamma_E + \ln 2}{m},
\end{equation}
where $\gamma_E$ is the Euler--Mascheroni constant~\cite{WhittakerWatson}. Substituting these values into the three-loop results of~\cite{Kazantsev2020,Haneychuk2022,Haneychuk2025} yields fully explicit $\beta$-functions parameterized by $(n,m)$, enabling a detailed study of scheme dependence and the role of finite coupling redefinitions in mapping between the $\overline{\text{DR}}$ and NSVZ schemes~\cite{Siegel1979,Jack1996_1,Jack1996_2}.

Finally, we note that explicit regulator-dependent structures are not merely technical artifacts: they illustrate how supersymmetric RG flows interpolate between different subtraction schemes, preserving NSVZ invariance under appropriate finite redefinitions. Moreover, they provide a natural starting point for connections to resurgent trans-series, anomaly matching, and holomorphic properties of supersymmetric gauge theories~\cite{Aniceto2021,Dunne2016,Marino2008,Anselmi1998}. Thus, evaluating $A(n)$ and $B(m)$ explicitly enriches our understanding of the interplay between exact RG structures, regularization, and scheme dependence in supersymmetric quantum field theory.

\medskip
The rest of this paper is organized as follows. In Sec.~\ref{sec:Preliminaries} we review the HCD setup, define notation, and recall the general structure of the three-loop $\beta$-function. In Sec.~\ref{sec:AB1} we evaluate $A(n)$ and $B(m)$ for exponential regulators, providing exact results and asymptotics. In Sec.~\ref{sec:ThreeLoopExponential} we substitute these constants into the known three-loop formulas and analyze scheme dependence. In Sec.~\ref{subsec:drbar} we compare explicitly with the compact general expression of Haneychuk~\cite{Haneychuk2025}. Finally, Sec.~\ref{sec:Conclusion} summarizes our results and discusses implications for NSVZ compatibility, scheme dependence, and the analytic structure of supersymmetric RG flows.

% --- Preliminaries ---
\section{Preliminaries}
\label{sec:Preliminaries}

This section collects the ingredients needed for computing multi–loop
renormalization group (RG) functions in $\mathcal{N}=1$ supersymmetric gauge
theories with higher covariant derivative (HCD) regularization supplemented by
Pauli–Villars (PV) superfields. We fix notation, summarize the gauge/matter
setup, recall the definitions of bare versus renormalized quantities, and
highlight the regulator–dependent constants $A$ and $B$ that will play a
central role in our three–loop analysis. Classic references on superspace and
conventions include~\cite{Gates1983,West1990,Buchbinder1998}; background on
multi–loop integrals and techniques can be found in~\cite{Smirnov2004}. The
HCD method goes back to Slavnov~\cite{Slavnov1971,Slavnov1972,Slavnov1977}
and, in the supersymmetric context, underlies modern diagrammatic derivations
of NSVZ–compatible RG relations and the structure of double total
derivatives~\cite{Kataev2013,Stepanyantz2020,Kazantsev2020,Haneychuk2022,Haneychuk2025}.

%----------------------------------------------------
\subsection{Conventions and notational choices}
\label{subsec:Conventions}
%----------------------------------------------------

We consider a semi–simple gauge group
\begin{equation}
\label{eq:gauge_group_def}
G \;=\; \prod_{K=1}^{n} G_K,
\end{equation}
where each $G_K$ is a simple Lie group ($SU(N)$, $SO(N)$, $Sp(N)$, \dots) or an
abelian factor $U(1)$. The corresponding vector superfields are denoted by
$V_K$. Chiral superfields $\{\Phi_a\}$ transform in representations $R_{aK}$
of $G_K$ (for $U(1)$, $R_{aK}$ reduces to a charge $q_{aK}$).

Bare gauge couplings are written as
\begin{equation}
\label{eq:alpha0_def}
\alpha_{0K}\;\equiv\;\frac{e_{0K}^{\,2}}{4\pi},
\end{equation}
and the corresponding renormalized couplings as
\begin{equation}
\label{eq:alpha_def}
\alpha_{K}\;\equiv\;\frac{e_{K}^{\,2}}{4\pi}.
\end{equation}
Bare Yukawa couplings are denoted collectively by $\lambda_0$ (with the index
structure specified below), while renormalized Yukawas are written without
subscript, $\lambda$.

Bare and renormalized objects are related by multiplicative renormalization
and, in general, finite (scheme-dependent) field and coupling redefinitions.
We define the \emph{bare} and \emph{renormalized} gauge $\beta$–functions by
\begin{equation}
\label{eq:beta_defs}
\beta_K(\alpha_0,\lambda_0)\;\equiv\;
\frac{d\,\alpha_{0K}}{d\ln\Lambda}\Big|_{\alpha,\lambda}\!,
\qquad
\widetilde{\beta}_K(\alpha,\lambda)\;\equiv\;
\frac{d\,\alpha_{K}}{d\ln\mu}\Big|_{\Lambda}\!,
\end{equation}
where $\Lambda$ is the UV cutoff (the HCD scale) and $\mu$ is the
renormalization scale. We reserve a tilde for renormalized RG functions
throughout.

The superpotential is normalized as
\begin{equation}
\label{eq:superpotential_norm}
W=\tfrac{1}{6}\,\lambda^{abc}\,\Phi_a\Phi_b\Phi_c,
\end{equation}
with completely symmetric $\lambda^{abc}$. We suppress gauge indices and use
trivial raising/lowering for flavor indices. We define the standard Yukawa
invariants
\begin{align}
\label{eq:yukawa_invariants}
(\lambda^\dagger\lambda)_a{}^b
&\equiv \lambda^*_{a c d}\,\lambda^{b c d},
&
(\lambda^\dagger C_K\lambda)_a{}^b
&\equiv
\lambda^*_{a c d}\,C(R_{cK})\,\lambda^{b c d},
\end{align}
and write repeated gauge–factor indices only when summed explicitly.

Dimensional reduction (DRED) and its minimal subtraction variant
$\overline{\mathrm{DR}}$ will be used as a reference subtraction scheme where
appropriate~\cite{Siegel1979,Jack1996_1,Jack1996_2}.

%----------------------------------------------------
\subsection{Gauge structure and group theory factors}
\label{subsec:GaugeStructure}
%----------------------------------------------------

We adopt standard group theory conventions:
\begin{equation}
\label{eq:group_conventions}
\mathrm{Tr}\!\left(T_{aK}^A T_{aK}^B\right) = T_{aK}\,\delta^{AB},\qquad
(T_{aK}^A T_{aK}^A)\,\phi_a = C(R_{aK})\,\phi_a,
\end{equation}
so that $T_{aK}$ is the Dynkin index of the representation $R_{aK}$ and
$C(R_{aK})$ its quadratic Casimir.

For abelian factors $U(1)$,
\begin{equation}
\label{eq:abelian_conventions}
T_{aK} = q_{aK}^2,\qquad C(R_{aK}) = q_{aK}^2.
\end{equation}

The adjoint Casimir is fixed by
\begin{equation}
\label{eq:casimir_def}
f^{ACD} f^{BCD} = C_2(G_K)\,\delta^{AB},
\end{equation}
with $f^{ABC}$ the structure constants of $G_K$.

It is convenient to introduce the one-loop combination
\begin{equation}
\label{eq:QK_def}
Q_K \;\equiv\; \sum_a T_{aK} - 3C_2(G_K),
\end{equation}
which coincides with the one-loop coefficient of the bare gauge $\beta$–function.

%----------------------------------------------------
\subsection{Renormalization and anomalous dimensions}
\label{subsec:RenormGamma}
%----------------------------------------------------

Bare and renormalized chiral superfields are related by a (generally
non–diagonal) wavefunction renormalization matrix $Z$,
\begin{equation}
\label{eq:field_renorm}
\Phi_a \;=\; \bigl(Z^{1/2}\bigr)_a{}^b\,\Phi_{Rb},
\end{equation}
where $Z_a{}^b$ depends on the UV cutoff $\Lambda$ and on the bare couplings
$(\alpha_{0K},\lambda_0)$.

The anomalous dimension matrix is defined by
\begin{equation}
\label{eq:gamma_def}
\gamma_a{}^b(\alpha_0,\lambda_0)
\;=\;
-\,\frac{d \ln Z_a{}^b}{d \ln \Lambda}\Big|_{\alpha,\lambda=\text{const}}.
\end{equation}
Two–loop expressions in supersymmetric gauge theories with Yukawas are
classical~\cite{Yamada1994}. In the HCD framework for semi–simple groups and
multiple couplings, compact formulas compatible with PV subtraction were
derived in Refs.~\cite{Kazantsev2020,Haneychuk2022,Haneychuk2025}. For later
use we quote the multi–coupling HCD form (see also Ref.~\cite{Korneev2021}):

\begin{widetext}
\begin{align}
\label{eq:gamma_HCD}
{\gamma_a}^b(\alpha_0,\lambda_0)
&= -\sum_{K} \frac{\alpha_{0K}}{\pi}\, C(R_{aK})\, {\delta_a}^b
+ \frac{1}{4\pi^2}\,(\lambda_{0}^\dagger \lambda_0)_a{}^b
+ \sum_{K,L}\frac{\alpha_{0K}\alpha_{0L}}{2\pi^2}
   C(R_{aK}) C(R_{aL})\, {\delta_a}^b
\nonumber\\[0.4ex]
&\quad - \sum_{K}\frac{\alpha_{0K}^2}{2\pi^2}C(R_{aK})
   \Bigl[\,3C_2(G_K)\ln a_{\varphi,K}
   - \sum_{c} T_{cK}\ln a_{K}
   - Q_K\!\Bigl(1+\tfrac{A}{2}\Bigr)\Bigr]\,{\delta_a}^b
\nonumber\\[0.4ex]
&\quad - \sum_{K}\frac{\alpha_{0K}}{8\pi^3}
   (\lambda_{0}^\dagger \lambda_0)_a{}^b\, C(R_{aK})\,(1 - B + A)
+ \sum_{K}\frac{\alpha_{0K}}{4\pi^3}
   (\lambda_{0}^\dagger C_K \lambda_0)_a{}^b\, (1 + B - A)
\nonumber\\[0.4ex]
&\quad - \frac{1}{16\pi^4}
   (\lambda_{0}^\dagger [\lambda_{0}^\dagger \lambda_0] \lambda_0)_a{}^b
   + \mathcal{O}\!\left(
     \alpha_0^3,\,
     \alpha_0^2 \lambda_0^2,\,
     \alpha_0 \lambda_0^4,\,
     \lambda_0^6\right).
\end{align}
\end{widetext}

Here $a_{\varphi,K}$ and $a_K$ are PV mass ratios (see
Sec.~\ref{subsec:HCD} below), and $A,B$ are regulator–dependent constants
defined in Sec.~\ref{subsec:HCD_finite}.

%----------------------------------------------------
\subsection{Bare and renormalized $\beta$–functions}
\label{subsec:BetaBareRenorm}
%----------------------------------------------------

The bare gauge $\beta$–function is defined in Eq.~\eqref{eq:beta_defs}.
Equivalently, for each gauge factor it is useful to keep the explicit form
\begin{equation}
\label{eq:beta_bare_detailed}
\beta_K(\alpha_0, \lambda_0)
\;=\;
\frac{d \alpha_{0K}}{d \ln \Lambda}\Big|_{\alpha,\lambda=\text{const}}.
\end{equation}

Bare RG functions depend on the regularization but are
\emph{subtraction-scheme independent}. This is precisely why they are the
natural objects for exact identities such as the NSVZ relation in the HCD+PV
framework. In contrast, the renormalized functions
$\widetilde{\beta}_K(\alpha,\lambda)$ depend on the subtraction prescription
(e.g.\ $\overline{\mathrm{DR}}$) and are related to the bare ones by finite,
analytic redefinitions of couplings and fields (see
Sec.~\ref{sec:FiniteRenorm})~\cite{Jack1996_1,Jack1996_2}.

%----------------------------------------------------
\subsection{NSVZ relation for bare couplings}
\label{subsec:NSVZbare}
%----------------------------------------------------

The Novikov–Shifman–Vainshtein–Zakharov (NSVZ) relation
\cite{Novikov1983,Shifman1986,Novikov1986,Jones1983,ArkaniHamed1997,Shifman1996,Seiberg1994}
provides an exact connection between gauge $\beta$–functions and anomalous
dimensions. In the HCD+PV framework, when RG functions are defined in terms of
\emph{bare} couplings, it takes the form~\cite{Stepanyantz2020}
\begin{align}
\label{eq:NSVZ_bare}
\frac{\beta_K(\alpha_0, \lambda_0)}{\alpha_{0K}^2}
&= -\frac{1}{2\pi \left(1 - \dfrac{C_2(G_K)\alpha_{0K}}{2\pi} \right)} \nonumber \\
&\quad \times \left[ 3C_2(G_K)
      - \sum_a T_{aK} \left(1 - \gamma_a{}^a(\alpha_0, \lambda_0) \right) \right].
\end{align}

Diagrammatically, a key structural feature of HCD is that multi–loop
integrands can be represented as (double) total derivatives in momentum space,
which is the mechanism behind the NSVZ structure and also explains why
regulator-dependent finite constants appear in compact higher-loop formulas
\cite{Kataev2013,Stepanyantz2020,Kazantsev2020}.

%----------------------------------------------------
\subsection{Higher covariant derivative regularization}
\label{subsec:HCD}
%----------------------------------------------------

The higher covariant derivative regularization modifies the classical action
by inserting gauge-covariant higher-derivative operators. In general, these
operators affect not only the quadratic part of the action, but also the
interaction vertices. For practical computations it is often convenient to
display explicitly only the quadratic part (which determines propagator
suppression in the ultraviolet) and to leave the vertex modifications implicit.

In particular, we write the regularized action in the schematic form
\begin{equation}
S^{\mathrm{reg}} = S_{\mathrm{cl}} + S_{\mathrm{HD}} + S_{\mathrm{gf}} + S_{\mathrm{ghost}} + S_{\mathrm{PV}},
\end{equation}
where $S_{\mathrm{HD}}$ contains the higher-derivative terms, $S_{\mathrm{gf}}$
is a gauge-fixing term compatible with supersymmetry, $S_{\mathrm{ghost}}$
contains the Faddeev–Popov and Nielsen–Kallosh ghosts, and $S_{\mathrm{PV}}$
implements the PV subtraction. Explicit superspace forms can be found, for
example, in Refs.~\cite{Slavnov1977,Kataev2013,Kazantsev2020}.

The purpose of the quadratic higher-derivative terms is to suppress ultraviolet
modes while preserving gauge invariance and supersymmetry
\cite{Slavnov1971,Slavnov1972}. The quadratic part of the regularized action
in the gauge and matter sectors takes the form
\begin{align}
\label{eq:gauge_reg_action}
S_{\text{gauge}}^{\text{reg}}
&= \frac{1}{2e_0^2}
\int d^4x\, d^4\theta \;
V \,
R\!\left(-\frac{\bar{D}^2 D^2}{16\Lambda^2}\right)
V ,
\\
\label{eq:matter_reg_action}
S_{\text{matter}}^{\text{reg}}
&= \frac{1}{4}
\int d^4x\, d^4\theta \;
\Phi^\dagger \,
F\!\left(-\frac{\bar{D}^2 D^2}{16\Lambda^2}\right)
\Phi .
\end{align}
Here $R(x)$ and $F(x)$ are regulator functions satisfying
\begin{equation}
\label{eq:RF_conditions}
R(0)=F(0)=1,
\qquad
R(x),\,F(x)\xrightarrow[x\to\infty]{}\infty,
\end{equation}
so that ultraviolet contributions are sufficiently damped.

After the introduction of higher derivatives, the remaining one-loop
divergences are eliminated by Pauli--Villars superfields with masses
proportional to the cutoff~$\Lambda$ \cite{Slavnov1977}. We parameterize the
corresponding PV mass ratios by $a_{\varphi,K}$ in the gauge sector and by
$a_K$ in the matter sector. These parameters are free regularization inputs
(subject to gauge invariance, supersymmetry, and decoupling conditions) and
affect only finite parts of multi-loop quantities; scheme-invariant
combinations remain unchanged (see also
Refs.~\cite{Jack1996_1,Jack1996_2,Korneev2021}).

%----------------------------------------------------
\subsection{Finite contributions from regulators and the constants
\texorpdfstring{$A$}{A}, \texorpdfstring{$B$}{B}}
\label{subsec:HCD_finite}
%----------------------------------------------------

Beyond one loop, HCD ensures overall UV finiteness of integrals but leaves
nontrivial finite remnants that depend on the explicit choice of regulator
functions. In the HCD formalism these finite pieces are universally encoded in
two constants~\cite{Kataev2013,Stepanyantz2020,Kazantsev2020,Haneychuk2022}:
\begin{align}
\label{eq:A_B_constants_multiline}
A &= \int_0^\infty dx\, \ln x\,
        \frac{d}{dx}\!\left(\frac{1}{R(x)}\right), \\
B &= \int_0^\infty dx\, \ln x\,
        \frac{d}{dx}\!\left(\frac{1}{F(x)^2}\right).
\end{align}
The constants $A$ and $B$ first appear in the two-loop anomalous dimensions,
see Eq.~\eqref{eq:gamma_HCD}, and enter crucially in general three-loop gauge
$\beta$–functions derived within HCD
\cite{Kazantsev2020,Haneychuk2022,Haneychuk2025}. Different admissible
regulator choices correspond to different values of $(A,B)$, reflecting the
finite scheme dependence of higher-loop coefficients in renormalized schemes
such as $\overline{\mathrm{DR}}$~\cite{Jack1996_1,Jack1996_2,Korneev2021}.

In this work we evaluate $A$ and $B$ explicitly for the exponential family
\begin{equation}
\label{eq:exp_reg_family}
R(x)=e^{x^n},\qquad F(x)=e^{x^m},\qquad n,m\ge 2,
\end{equation}
which affords analytic control and strong UV suppression. The results,
\begin{equation}
\label{eq:AB_exp_results}
A(n)=\frac{\gamma_E}{n},\qquad B(m)=\frac{\gamma_E+\ln 2}{m},
\end{equation}
with $\gamma_E$ the Euler–Mascheroni constant~\cite{WhittakerWatson}, will be
derived in Sec.~\ref{sec:AB1} and inserted into the compact three-loop HCD
expressions of Refs.~\cite{Kazantsev2020,Haneychuk2022,Haneychuk2025} in
Sec.~\ref{sec:ThreeLoopExponential}. This will allow us to track precisely how
finite regulator-tagged pieces are reshuffled by finite redefinitions when
mapping between NSVZ–compatible bare expressions and renormalized schemes such
as $\overline{\mathrm{DR}}$~\cite{Siegel1979,Jack1996_1,Jack1996_2}.

\paragraph*{Pauli--Villars masses.}
Within HCD, the PV superfields cancel the remaining one-loop divergences. The
ratios $a_{\varphi,K}$ and $a_K$ are free (admissible) inputs of the
regularization and shift only finite parts of multi-loop coefficients. In
particular, they do not affect scheme-invariant combinations entering
multi-loop RG functions, and different admissible choices may be used to
simplify intermediate expressions without changing physical content
\cite{Slavnov1977,Kataev2013,Korneev2021}.

\medskip
For completeness, we also recall that the appearance of (double) total
derivatives in momentum space is a characteristic feature of HCD that
facilitates both the derivation of NSVZ-type relations and the isolation of
the constants $A$ and $B$ \cite{Kataev2013,Stepanyantz2020}. In what follows we
use this mechanism only through the compact results quoted above and do not
rederive it here.

%====================================================
\section{Explicit Computation of the Regulator-Dependent Constants $A$ and $B$}
\label{sec:AB1}
%====================================================

In the higher covariant derivative (HCD) framework, finite
regulator-dependent contributions to multi-loop quantities are
universally encoded in two constants, $A$ and $B$. These constants
are constructed from the gauge- and matter-sector regulator
functions $R(x)$ and $F(x)$ that modify the quadratic part of the
regularized action (see Sec.~\ref{subsec:HCD})
\cite{Slavnov1971,Slavnov1972,Slavnov1977,Kataev2013,Stepanyantz2020}.
In the compact three-loop expressions derived within HCD for
semi-simple gauge groups with Yukawa interactions
\cite{Kazantsev2020,Haneychuk2022,Haneychuk2025},
$A$ and $B$ appear as finite parameters controlled entirely by the
choice of regulator functions. Their explicit evaluation for a given
regulator family is therefore essential for obtaining concrete
three-loop $\beta$-functions and for organizing scheme dependence.

%----------------------------------------------------
\subsection{Definitions and convergence}
\label{subsec:AB_def_conv}
%----------------------------------------------------

The constants $A$ and $B$ are defined by
\begin{align}
\label{eq:AB_def}
A &\equiv \int_0^\infty dx \,\ln x\,
      \frac{d}{dx}\!\left(\frac{1}{R(x)}\right),
\\
B &\equiv \int_0^\infty dx \,\ln x\,
      \frac{d}{dx}\!\left(\frac{1}{F(x)^2}\right).
\end{align}

Admissible regulator functions satisfy
\[
R(0)=F(0)=1,
\qquad
R(x),F(x)\xrightarrow[x\to\infty]{}\infty.
\]
For such regulators the large-$x$ region is exponentially suppressed,
ensuring ultraviolet convergence of the integrals in
\eqref{eq:AB_def}.

Near $x\to0$, the logarithmic weight $\ln x$ can potentially expose
integrable singularities. However, once the explicit exponential
regulators are substituted, the integrands behave as $x^{n-1}\ln x$
or $x^{m-1}\ln x$, which are integrable at the origin for
$n,m\ge2$. Consequently, no boundary terms arise and the integrals
are finite without further regularization.

For completeness, one may verify the finite parts by Mellin analytic
continuation; this provides a useful cross-check but is not required
for the elementary derivations below.

%----------------------------------------------------
\subsection{Mellin building block (cross-check)}
\label{subsec:AB_mellin}
%----------------------------------------------------

A frequently used identity is
\begin{equation}
\label{eq:mellin_basic}
\int_0^\infty x^{s-1} e^{-x^p}\,dx
=
\frac{1}{p}\,
\Gamma\!\left(\frac{s}{p}\right).
\end{equation}
Expanding around $s\to0$,
\[
\Gamma\!\left(\frac{s}{p}\right)
=
\frac{p}{s}
-
\gamma_E
+
\mathcal{O}(s),
\]
one obtains the finite part
\begin{equation}
\label{eq:IR_block}
\mathrm{FP}
\left\{
\int_0^\infty \frac{dx}{x}\,e^{-x^p}
\right\}
=
-\frac{\gamma_E}{p},
\end{equation}
where $\gamma_E$ is the Euler–Mascheroni constant.
This identity will serve only as a consistency check.

%----------------------------------------------------
\subsection{Evaluation of $A(n)$ for exponential gauge regulators}
\label{subsec:A_eval}
%----------------------------------------------------

Consider the exponential gauge-sector family
\begin{equation}
\label{eq:R_exp}
R(x)=e^{x^n},
\qquad n\ge2.
\end{equation}
Then
\[
\frac{1}{R(x)}=e^{-x^n},
\qquad
\frac{d}{dx}\!\left(\frac{1}{R(x)}\right)
=
- n x^{n-1} e^{-x^n}.
\]

Substituting into \eqref{eq:AB_def},
\begin{align}
A(n)
&=
- n \int_0^\infty x^{n-1}\ln x\,e^{-x^n}\,dx.
\end{align}

With the change of variables
\[
t=x^n,
\qquad
dt=n x^{n-1}dx,
\qquad
\ln x=\frac{1}{n}\ln t,
\]
we obtain
\begin{align}
A(n)
&=
- \int_0^\infty
\frac{1}{n}\ln t\,e^{-t}\,dt.
\end{align}

Using the standard integral
\[
\int_0^\infty e^{-t}\ln t\,dt=-\gamma_E,
\]
we find
\begin{equation}
\label{eq:A_result}
A(n)=\frac{\gamma_E}{n}.
\end{equation}

The result is finite and no endpoint contributions arise for $n\ge2$.

%----------------------------------------------------
\subsection{Evaluation of $B(m)$ for exponential matter regulators}
\label{subsec:B_eval}
%----------------------------------------------------

For the exponential matter-sector family
\begin{equation}
\label{eq:F_exp}
F(x)=e^{x^m},
\qquad m\ge2,
\end{equation}
we have
\[
\frac{1}{F(x)^2}=e^{-2x^m},
\qquad
\frac{d}{dx}\!\left(\frac{1}{F(x)^2}\right)
=
-2m x^{m-1} e^{-2x^m}.
\]

Hence
\begin{align}
B(m)
&=
-2m \int_0^\infty x^{m-1}\ln x\,e^{-2x^m}\,dx.
\end{align}

Introduce the change of variables
\[
u=2x^m,
\qquad
du=2m x^{m-1}dx,
\qquad
\ln x=\frac{1}{m}(\ln u-\ln2).
\]
Then
\begin{align}
B(m)
&=
-\int_0^\infty
\left(
\frac{1}{m}\ln u
-
\frac{\ln2}{m}
\right)
e^{-u}\,du.
\end{align}

Using
\[
\int_0^\infty e^{-u}\ln u\,du=-\gamma_E,
\qquad
\int_0^\infty e^{-u}\,du=1,
\]
we obtain
\begin{equation}
\label{eq:B_result}
B(m)=\frac{\gamma_E+\ln2}{m}.
\end{equation}

Again the integral is finite for $m\ge2$.

%----------------------------------------------------
\subsection{Scaled exponentials}
\label{subsec:scaled}
%----------------------------------------------------

For completeness consider the scaled family
\begin{equation}
R(x)=e^{c x^p},
\qquad
F(x)=e^{c x^p},
\qquad
c>0,\ p\ge2.
\end{equation}
Using the Mellin representation \eqref{eq:mellin_basic} one finds
\begin{equation}
\mathrm{FP}
\left\{
\int_0^\infty \frac{dx}{x}\,e^{-c x^p}
\right\}
=
-\frac{\gamma_E+\ln c}{p}.
\end{equation}

Applying the definitions in \eqref{eq:AB_def} yields
\begin{equation}
\label{eq:A_B_scaled}
A(p;c)=\frac{\gamma_E+\ln c}{p},
\qquad
B(p;c)=\frac{\gamma_E+\ln(2c)}{p}.
\end{equation}

In what follows we mostly use $c=1$.

For the regulator choice \eqref{eq:R_exp} and \eqref{eq:F_exp} we
combine \eqref{eq:A_result} and \eqref{eq:B_result} into
\begin{equation}
\label{eq:AB_results_final}
A(n)=\frac{\gamma_E}{n},
\qquad
B(m)=\frac{\gamma_E+\ln2}{m}.
\end{equation}

These closed-form expressions will be substituted into the compact
three-loop HCD results of
\cite{Kazantsev2020,Haneychuk2022,Haneychuk2025}
in Sec.~\ref{sec:ThreeLoopExponential}.
They make explicit how regulator-dependent finite pieces enter
the three-loop gauge $\beta$-functions and allow a transparent
analysis of how such pieces are redistributed by finite
redefinitions when mapping to renormalized schemes such as
$\overline{\mathrm{DR}}$
\cite{Siegel1979,Jack1996_1,Jack1996_2}.

% --- Three‐Loop $\beta$-Functions with Exponential Regulators ---
\section{Three‐Loop \texorpdfstring{$\beta$}{$\beta$}-Functions with Exponential Regulators}
\label{sec:ThreeLoopExponential}

In this section, we \emph{obtain explicit three-loop expressions by substituting}
the evaluated regulator parameters \(A(n)\) and \(B(m)\) for exponential regulators into the
\emph{previously derived} general HCD three-loop formulas. We do not re-derive the general
three-loop expressions; instead we make their dependence on the exponential regulator family
\(R(x)=e^{x^n}\), \(F(x)=e^{x^m}\) manifest, and then compare the resulting bare and renormalized
\(\beta\)-functions (structure-by-structure) with the compact formulas reported in
Ref.~\cite{Haneychuk2022, Haneychuk2025}.

\subsection{Framework, Regulators, and PV Masses}
\label{subsec:framework_regulator}

We consider a semi-simple gauge group
\begin{equation}
  G \;=\; \prod_{K} G_{K},
\end{equation}
with gauge couplings \(g_K\) (we frequently use \(\alpha_{K}\equiv g_K^2/(4\pi)\)). Chiral superfields \(\phi_a\)
transform in representations \(R_{aK}\) of \(G_K\), and Yukawa interactions are encoded in
\begin{equation}
  W \;=\; \tfrac{1}{6}\,\lambda^{abc}\,\phi_a \phi_b \phi_c \,,
\end{equation}
with \(\lambda^{abc}\) totally symmetric in its flavor indices.

We employ higher covariant derivative (HCD) regularization supplemented by Pauli--Villars (PV)
superfields. The regulator functions are chosen to be exponential:
\begin{align}
R(x) &= e^{x^{n}}, \qquad F(x) = e^{x^{m}}, \notag \\
x &\equiv \frac{p^{2}}{\Lambda^{2}}, \qquad n,m \in \mathbb{N}, \quad n,m \ge 2.
\end{align}
These satisfy the standard HCD admissibility conditions
\cite{Slavnov1971,Slavnov1972,Slavnov1977,Kataev2013,Stepanyantz2020}:
\begin{align}
  R(0) &= F(0) = 1, \\
  R(x),\, F(x) &> 0 \quad \text{and monotone for } x \ge 0, \\
  \lim_{x \to \infty} R(x),\, F(x) &= +\infty \quad \text{(sufficient UV growth)}.
\end{align}
In this setup the two regulator-dependent constants entering the three-loop HCD formulae are defined
by convergent integrals (see Appendix~\ref{app:RegulatorDetails} for details and
divergence-cancellation steps)
\begin{equation}
\begin{split}
A &\equiv \int_{0}^{\infty} dx \,\ln x\,
      \frac{d}{dx}\!\left(\frac{1}{R(x)}\right), \\
B &\equiv \int_{0}^{\infty} dx \,\ln x\,
      \frac{d}{dx}\!\left(\frac{1}{F(x)^{2}}\right).
\end{split}
\end{equation}
and for the exponential family we obtain in closed form
\begin{equation}
  A(n)=\frac{\gamma_E}{n}\,,\qquad B(m)=\frac{\gamma_E+\ln 2}{m}\,,
  \label{eq:A_B_values}
\end{equation}
where \(\gamma_E\) is the Euler--Mascheroni constant. We stress that \emph{PV masses are free
regularization parameters}, subject to gauge invariance, supersymmetry, and decoupling constraints
\cite{Slavnov1977,Kataev2013}. Our particular PV spectrum is chosen for technical convenience and is
justified in Secs.~\ref{sec:AB1}--\ref{sec:ThreeLoopExponential}; it affects only \emph{finite, scheme-dependent}
pieces and leaves scheme-invariant multi-loop structures unchanged.

\subsection{Comparison: exponential vs.\ polynomial-type HCD regulators}
\label{subsec:RegComp}

While this work focuses on the exponential family $R(x)=e^{x^n}$, $F(x)=e^{x^m}$, it is instructive to contrast
the structure of the regulator-dependent constants with common polynomial-type choices (e.g.\
$R_{\text{poly}}(x)=(1+x^{p})^{\rho}$, $F_{\text{poly}}(x)=(1+x^{q})^{\sigma}$) used in the HCD literature
(see, e.g., \cite{Kataev2013,Stepanyantz2020,Korneev2021}). In all admissible cases the finite constants are captured
by the same master definitions,
\[
A \;=\; \int_0^\infty dx\,\ln x\,\frac{d}{dx}\frac{1}{R(x)}\!,\qquad
B \;=\; \int_0^\infty dx\,\ln x\,\frac{d}{dx}\frac{1}{F(x)^2}\!,
\]
but closed forms depend on the explicit profile of $R,F$. For the profiles indicated above one finds the following
universal $1/\text{(power)}$ scaling:
\begin{table}[h]
\centering
\begin{tabular}{@{}lcc@{}}
\toprule
\textbf{Regulator family} & \boldmath$A$ (gauge) & \boldmath$B$ (matter) \\
\midrule
Exponential: $R = e^{x^n}$,\quad $F = e^{x^m}$
    & $A(n) = \dfrac{\gamma_E}{n}$
    & $B(m) = \dfrac{\gamma_E + \ln 2}{m}$ \\
Polynomial-type: $R = (1 + x^p)^\rho$
    & $A \sim \dfrac{c_R(\rho)}{p}$
    & --- \\
Polynomial-type: $F = (1 + x^q)^\sigma$
    & ---
    & $B \sim \dfrac{c_F(\sigma)}{q}$ \\
\bottomrule
\end{tabular}
\caption{Asymptotic forms of the constants $A$ and $B$ for different regulator families; $c_R(\rho)$ and $c_F(\sigma)$ are finite
constants that tag the scheme choice and vanish in the large-power limit.}
\label{tab:A_B_constants}
\end{table}

\subsection{NSVZ Structure for Bare Couplings and Notation}
\label{subsec:nsvz_bare_notation}

Defining \(\alpha_{0K}\equiv g_{0K}^2/(4\pi)\) and \(\lambda_{0}\) as bare couplings, the NSVZ form of the
\emph{bare} gauge \(\beta\)-functions in HCD reads \cite{Novikov1983,Shifman1986,Novikov1986,Jones1983,Stepanyantz2020}
\begin{align}
\label{eq:NSVZ_three_loop}
\frac{\beta_K(\alpha_0,\lambda_0)}{\alpha_{0K}^{2}}
&= -\frac{1}{2\pi\Bigl(1-\tfrac{C_2(G_K)\,\alpha_{0K}}{2\pi}\Bigr)} \notag \\
&\quad \times \biggl[\,3\,C_2(G_K)\;-\;\sum_{a} T_{aK}\Bigl(1-\gamma^{a}{}_{a}(\alpha_0,\lambda_0)\Bigr)\biggr],
\end{align}
with \(\gamma^{a}{}_{a}\) the (matrix) anomalous dimension of chiral fields. We use group-theory conventions
\begin{align}
(T^{A}T^{A})_{R} &= C(R)\,\mathbf{1}, \notag \\
f^{ACD}f^{BCD} &= C_2(G)\,\delta^{AB}, \notag \\
Q_{K} &\equiv \sum_{a} T_{aK}-3\,C_2(G_K)\,,
\end{align}
where \(Q_{K}\) coincides with the one-loop coefficient of the bare \(\beta\)-function.

\subsection{Three-Loop Bare \texorpdfstring{$\beta$}{beta}-Function with Exponential Regulators}
\label{subsec:barebeta}

Substituting the two-loop anomalous dimensions \(\gamma^{a}{}_{a}(\alpha_0,\lambda_0)\) into
Eq.~\eqref{eq:NSVZ_three_loop}, and keeping Yukawas explicitly, we obtain for each gauge factor \(G_K\)
the three-loop \emph{bare} result (see also \cite{Kazantsev2020,Haneychuk2022,Haneychuk2025}):
\begin{widetext}
\begin{align}
\label{eq:beta_bare_full}
\frac{\beta_K(\alpha_0,\lambda_0)}{\alpha_{0K}^{2}}
= -\frac{1}{2\pi}\Biggl\{ &
-\,Q_K - \frac{\alpha_{0K}}{2\pi}\,C_2(G_K)\,Q_K
- \sum_{a,L}\frac{\alpha_{0L}}{\pi}\,T_{aK}\,C(R_{aL}) \notag \\
&+ \frac{1}{4\pi^2}\sum_{abc} T_{aK}\,\lambda_0^{\dagger abc}\lambda_{0\,abc}
- \sum_{a,L}\frac{\alpha_{0K}\alpha_{0L}}{2\pi^2}\,T_{aK}\,C_2(G_K)\,C(R_{aL}) \notag \\
&- \frac{\alpha_{0K}^2}{4\pi^2}\,C_2^{2}(G_K)\,Q_K
+ \sum_{a,M,N}\frac{\alpha_{0M}\alpha_{0N}}{2\pi^2}\,T_{aK}\,C(R_{aM})\,C(R_{aN}) \notag \\
&- \sum_{a,L}\frac{\alpha_{0L}^2}{2\pi^2}\,T_{aK}\,C(R_{aL}) \Bigl[\,3\,C_2(G_L)\ln a_{\varphi,L} - \sum_b T_{bL}\ln a_L - Q_L\Bigl(1+\tfrac{A(n)}{2}\Bigr)\Bigr] \notag \\
&- \sum_{abc,L}\frac{\alpha_{0L}}{8\pi^3}\,T_{aK}\,C(R_{aL})\,\lambda_0^{\dagger abc}\lambda_{0\,abc} \Bigl(\,1 + A(n) - B(m)\Bigr) \notag \\
&+ \sum_{abc,L}\frac{\alpha_{0L}}{4\pi^3}\,T_{aK}\,\lambda_0^{\dagger abc}\,C(R_{cL})\,\lambda_{0\,abc} \Bigl(\,1 + B(m) - A(n)\Bigr) \notag \\
&+ \sum_{abc}\frac{\alpha_{0K}}{8\pi^3}\,T_{aK}\,C_2(G_K)\,\lambda_0^{\dagger abc}\lambda_{0\,abc} \notag \\
&- \frac{1}{16\pi^4}\sum_{abcdef} T_{aK}\,\lambda_0^{\dagger abc}\lambda_{0\,cde}\lambda_0^{\dagger def}\lambda_{0\,abf}
\Biggr\} \,+\, \mathcal{O}(\alpha_0^3).
\end{align}
\end{widetext}

\noindent\textit{Comments.}
(i) The constants \(A,B\) originate from convergent linear combinations of loop integrals; all intermediate
singular contributions cancel analytically (see Appendix~\ref{app:RegulatorDetails}).
(ii) In the mixed gauge--Yukawa sector, the \emph{difference} \(A-B\) (or \(B-A\)) appears, rather than any
ratio \(B/A\); this exactly matches the general HCD formula
\cite{Kazantsev2020,Haneychuk2022,Haneychuk2025}.
(iii) PV mass parameters influence only the \emph{finite} parts; none of the scheme-invariant combinations
are affected by admissible PV choices \cite{Slavnov1977,Kataev2013}.

\subsection{Finite Redefinitions and Scheme Mapping}
\label{subsec:finite_map}

The relation between bare and renormalized couplings is accompanied by admissible \emph{finite}
redefinitions. For multiple gauge factors and Yukawas, up to the order relevant for three-loop gauge
\(\beta\)-functions:
\begin{align}
\alpha_{K}' \,&=\, \alpha_{K}
  \;+\; \sum_{L} r^{(1)}_{KL}\,\alpha_{K}\alpha_{L}
  \;+\; \sum_{L,M} r^{(2)}_{KLM}\,\alpha_{K}\alpha_{L}\alpha_{M} \notag \\
  &\quad \;+\; s^{(1)}_{K\,abc}\,\alpha_{K}\,\lambda_{abc}\lambda^{abc}
  \;+\; \mathcal{O}(\alpha^{4},\alpha^{2}\lambda^{2}) \,, 
\label{eq:finite_map_alpha}
\\[2pt]
\lambda'_{abc} \,&=\, \lambda_{abc}
  \;+\; u^{(1)}_{abc,K}\,\alpha_{K}\lambda_{abc}
  \;+\; \mathcal{O}(\alpha^{2}\lambda,\lambda^{3}) \,,
\label{eq:finite_map_lambda}
\end{align}
with constant tensors \(r^{(1)},r^{(2)},s^{(1)},u^{(1)}\) encoding the finite parts
\cite{Jack1996_1,Jack1996_2}. Under Eqs.~\eqref{eq:finite_map_alpha}--\eqref{eq:finite_map_lambda},
the renormalized \(\beta\)-functions transform as
\begin{equation}
  \beta'_{K}(\alpha',\lambda') \;=\;
  \sum_{L}\frac{\partial \alpha_{K}'}{\partial \alpha_{L}}\,\beta_{L}(\alpha,\lambda)
  \;+\;
  \sum_{a,b,c}\frac{\partial \alpha_{K}'}{\partial \lambda_{abc}}\,\beta_{\lambda_{abc}}(\alpha,\lambda)\,,
\end{equation}
so that purely \emph{finite} constants can be shifted between schemes without altering any
scheme-invariant content \cite{Siegel1979,Shifman1996,Stepanyantz2020}. This observation will be used below
to align renormalized results.

\subsection{Renormalized \texorpdfstring{$\widetilde{\beta}_K$}{$\beta$-K} and Scheme Dependence}
\label{subsec:renorm_beta}

We relate bare and renormalized couplings (at scale \(\mu\)) by
\begin{equation}
\label{eq:bare_renorm_relation}
  \frac{1}{\alpha_{0K}} \;=\; \frac{1}{\alpha_K} \;+\; \frac{1}{2\pi}
  \Bigl(Q_K\ln\tfrac{\Lambda}{\mu} - b_{1,K}\Bigr) \;+\; \mathcal{O}(\alpha)\,,
\end{equation}
where \(b_{1,K}\) and higher constants capture \emph{finite} scheme dependence.
Using the chain rule with Eq.~\eqref{eq:beta_bare_full}, we obtain the renormalized three-loop
\(\beta\)-function (cf.\ \cite{Kazantsev2020,Haneychuk2022}):
\begin{widetext}
\begin{align}
\label{eq:BetaRenormalizedFinal}
\frac{\widetilde{\beta}_K(\alpha)}{\alpha_K^2} \;=\; -\frac{1}{2\pi}\Biggl\{ &
-\,Q_K
- \frac{\alpha_K}{2\pi}\,C_2(G_K)\,Q_K
- \sum_{a,L}\frac{\alpha_L}{\pi}\,T_{aK}\,C(R_{aL}) \notag \\
&\quad - \sum_{a,L}\frac{\alpha_K\alpha_L}{2\pi^2}\,T_{aK}\,C_2(G_K)\,C(R_{aL})
- \frac{\alpha_K^2}{4\pi^2}\,C_2(G_K)\,Q_K\Bigl(C_2(G_K)+b_{2,K}-b_{1,K}\Bigr) \notag \\
&\quad + \sum_{a,M,N}\frac{\alpha_M\alpha_N}{2\pi^2}\,T_{aK}\,C(R_{aM})\,C(R_{aN}) \notag \\
&\quad - \sum_{a,L}\frac{\alpha_L^2}{2\pi^2}\,T_{aK}\,C(R_{aL}) \Bigl[\,3\,C_2(G_L)\ln a_{\varphi,L} - \sum_b T_{bL}\ln a_L - b_{1,L} \notag \\
&\quad - Q_L\Bigl(1+b_{2,KL}+\tfrac{A(n)}{2}\Bigr)\Bigr]
\Biggr\} \;+\; \mathcal{O}(\alpha^3).
\end{align}
\end{widetext}

\subsection{Specialization to \texorpdfstring{$\overline{\mathrm{DR}}$}{DR\textendash bar}}
\label{subsec:drbar}

\begin{figure*}[ht]
  \centering
  \includegraphics[width=\textwidth]{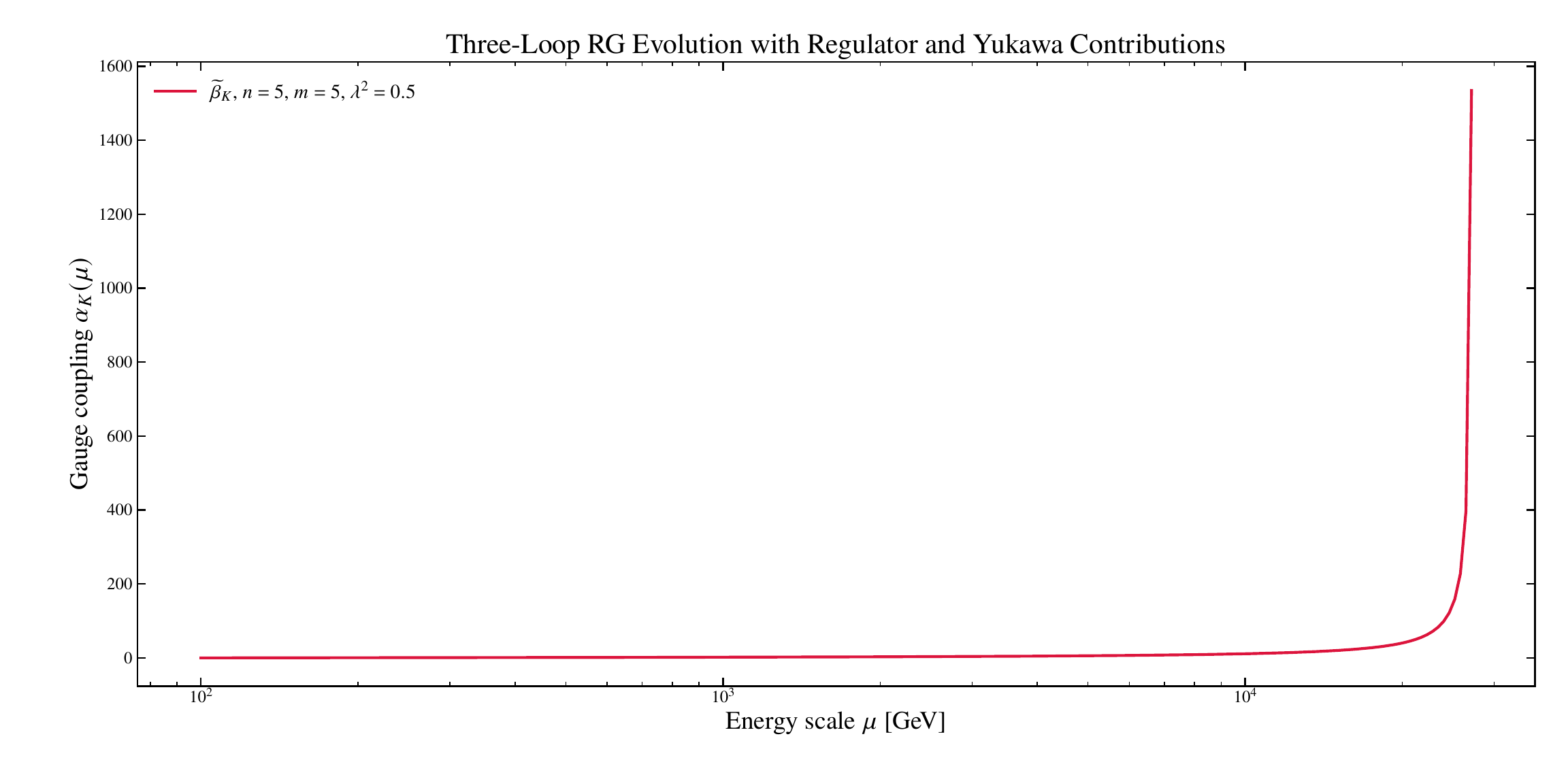}
  \caption{
    Illustrative RG evolution of \(\alpha_K(\mu)\) governed by the full three-loop
    \(\widetilde{\beta}_K\), including gauge and Yukawa contributions and the regulator-dependent terms.
    The running is computed in \(\overline{\mathrm{DR}}\) with \((n,m)=(5,5)\) and a sample Yukawa strength
    \(\lambda^2=0.5\). Parameters are \emph{illustrative}; the plot demonstrates sensitivity to finite,
    scheme-dependent pieces rather than making phenomenological claims.
  }
  \label{fig:RG_Yukawa_Regulator}
\end{figure*}

For the \(\overline{\mathrm{DR}}\) scheme (minimal subtraction in DRED), it is convenient to parametrize
the finite parts of the gauge-coupling renormalization by the constants \(b_{1,K}\), \(b_{2,K}\), and
\(b_{2,KL}\), defined so that the relation between bare and renormalized couplings is
\[
  \frac{1}{\alpha_{0K}}
  \;=\;
  \frac{1}{\alpha_{K}}
  + \frac{1}{2\pi}\!\left(Q_{K}\ln\frac{\Lambda}{\mu}-b_{1,K}\right)
  + \mathcal{O}(\alpha)\,,
\]
and the three–loop renormalized \(\widetilde{\beta}\)-function can be written in terms of these constants
(see Eq.~\eqref{eq:BetaRenormalizedFinal}). In \(\overline{\mathrm{DR}}\) one finds
\begin{align}
\label{eq:bDR_definitions}
  b_{1,K} &= 3\,C_2(G_K)\ln a_{\varphi,K} - \sum_{a} T_{aK}\ln a_K, \\
  b_{2,K} &= b_{1,K} - \tfrac{1}{4}Q_K, \\
  b_{2,KL} &= -\tfrac{1}{4} - \tfrac{A(n)}{2},
\end{align}
where the first line organizes the finite parts tied to gauge and matter wavefunction normalizations
through the PV mass ratios \(a_{\varphi,K}\) and \(a_{K}\); the second line isolates the universal
\(-\tfrac{1}{4}Q_{K}\) piece characteristic of minimal subtraction–type schemes; and the mixed
two–coupling constant \(b_{2,KL}\) contains both the MS–like \(-\tfrac{1}{4}\) and the regulator
dependent \(-\tfrac{A}{2}\) contribution. Using the exponential regulator family, \(A(n)=\gamma_{E}/n\)
[Eq.~\eqref{eq:A_B_values}], this becomes \(b_{2,KL}=-\tfrac{1}{4}-\tfrac{\gamma_{E}}{2n}\).

\medskip

The constants in \eqref{eq:bDR_definitions} encode how finite subtractions redistribute
three–loop contributions between different tensor structures once the RG is expressed in terms of
\emph{renormalized} couplings. In DRED, disappearing degrees of freedom only affect finite parts in
supersymmetric gauge theories, so the scheme dependence relevant here can be captured completely by
\(b_{1,K}\), \(b_{2,K}\), and \(b_{2,KL}\) \cite{Siegel1979,Jack1996_1,Jack1996_2}.
The mixed coefficient \(b_{2,KL}\) is particularly informative: its universal term \(-\tfrac{1}{4}\) is the
same for all admissible HCD regulators, while the additional shift \(-\tfrac{A}{2}\) remembers the
regulator choice made when computing the bare amplitudes (here through \(A(n)\)).
In the limit \(n\to\infty\) one has \(A(n)\to 0\), and \(b_{2,KL}\) approaches its MS–like value
\(-\tfrac{1}{4}\).

\medskip

Starting from the HCD/NSVZ bare form and the identity
\(
  \beta_{K}(\alpha_{0},\lambda_{0})
  =
  \frac{d\alpha_{0K}}{d\ln\Lambda}\Big|_{\alpha,\lambda}
\),
write \(1/\alpha_{0K}\) as above and expand to three loops keeping all terms up to
\(\mathcal{O}(\alpha^{2})\) in the curly braces of Eq.~\eqref{eq:BetaRenormalizedFinal}.
Matching coefficients of the independent group–theory tensors and Yukawa invariants then fixes
\(b_{1,K}\), \(b_{2,K}\), and \(b_{2,KL}\) uniquely in terms of the finite parts present in the bare
combinations. The appearance of \(A/2\) in \(b_{2,KL}\) follows from the pieces generated by differentiating
the finite (regulator–dependent) terms with respect to \(\ln\mu\) when converting \(\alpha_{0}\) to \(\alpha\).

\medskip

\noindent\textbf{NSVZ restoration by finite redefinitions.}
Because \(\overline{\mathrm{DR}}\) collects all nonlogarithmic remnants into \(b\)-constants, it does not
display the NSVZ form manifestly beyond two loops \cite{Jack1996_1,Jack1996_2}. Nevertheless, a finite scheme transformation of the
form
\[
  \alpha'_{K}
  \;=\;
  \alpha_{K}
  + \sum_{L} r^{(1)}_{KL}\,\alpha_{K}\alpha_{L}
  + \sum_{L,M} r^{(2)}_{KLM}\,\alpha_{K}\alpha_{L}\alpha_{M}
  + \cdots,
\]
possibly accompanied by a corresponding finite redefinition of \(\lambda\), maps the DR–bar result into an
NSVZ–compatible scheme. Choosing \(r^{(1)}\) and \(r^{(2)}\) to cancel the finite remnants proportional to
\(b_{2,KL}+\tfrac{A}{2}\) restores the NSVZ denominator structure
\(\bigl(1-\tfrac{C_{2}(G_{K})\alpha_{K}}{2\pi}\bigr)^{-1}\) in the renormalized \(\beta\)–function. Since the
three–loop scheme invariants are unaffected by such finite redefinitions, physical conclusions derived
from scheme–independent combinations remain unchanged.

\medskip

With these constants fixed, the renormalized three–loop \(\beta\)–function in \(\overline{\mathrm{DR}}\) takes
the compact form
\begin{widetext}
\begin{align}
\label{eq:BetaDR}
\left.\frac{\widetilde{\beta}_K(\alpha)}{\alpha_K^2}\right|_{\overline{\mathrm{DR}}}
= -\frac{1}{2\pi}\Biggl\{\, 
  -\,Q_K 
  - \frac{\alpha_K}{2\pi}\,C_2(G_K)\,Q_K 
  - \sum_{a,L}\frac{\alpha_L}{\pi}\,T_{aK}\,C(R_{aL})
  - \sum_{a,L}\frac{\alpha_K\alpha_L}{2\pi^2}\,T_{aK}\,C_2(G_K)\,C(R_{aL})  \notag\\
  \qquad\qquad
  - \frac{\alpha_K^2}{4\pi^2}\,C_2(G_K)\,Q_K\Bigl(C_2(G_K)-\tfrac{1}{4}Q_K\Bigr)
  + \sum_{a,M,N}\frac{\alpha_M\alpha_N}{2\pi^2}\,T_{aK}\,C(R_{aM})\,C(R_{aN})
  + \sum_{a,L}\frac{3\,\alpha_L^2}{8\pi^2}\,T_{aK}\,C(R_{aL})\,Q_L 
\Biggr\} \;+\; \mathcal{O}(\alpha^3).
\end{align}
\end{widetext}
Equations~\eqref{eq:bDR_definitions}–\eqref{eq:BetaDR} are consistent with the general HCD identity
\(b_{2,KL}=-\tfrac{1}{4}-\tfrac{A}{2}\); for the exponential family \(A(n)=\gamma_E/n\), in agreement with
Eq.~\eqref{eq:A_B_values}. The limit \(n\to\infty\) suppresses regulator–dependent finite parts and
smoothly connects back to the MS–like mixed coefficient \(b_{2,KL}=-\tfrac{1}{4}\), providing a useful
cross–check of the construction.

\medskip

To compare with the compact general HCD formulas of Ref.~\cite{Haneychuk2025}, we adopt the
following dictionary (hats denote the conventions of Ref.~\cite{Haneychuk2025}):
\begin{equation}
\begin{aligned}
  \widehat{\alpha}_{K} &\;\widehat{=}\; \alpha_{K}, \\
  \{\,\widehat{C}_{2}(G_{K}),\,\widehat{T}_{aK},\,\widehat{C}(R_{a})\,\}
  &\;\widehat{=}\; \{\,C_{2}(G_{K}),\,T_{aK},\,C(R_{a})\,\}, \\
  \widehat{A} &\;\widehat{=}\; A(n), \qquad
  \widehat{B} \;\widehat{=}\; B(m).
\end{aligned}
\end{equation}
Any difference in the matter-sector regulator power (\(F\) versus \(F^{2}\)) is absorbed into a redefinition
of \(B\) (i.e.\ \(B\to\tilde{B}\)) via a trivial rescaling of the defining integral.\footnote{Our definition
of \(B\) uses \(F(x)^{2}\) in the denominator, which is common in the HCD literature; if a different power
is employed elsewhere, \(\tilde{B}=B+\text{const.}\) modifies only finite terms.}

With this dictionary in place and after inserting \(A(n)=\gamma_{E}/n\) and \(B(m)=(\gamma_{E}+\ln 2)/m\),
the \emph{bare} result in Eq.~\eqref{eq:beta_bare_full} agrees structure-by-structure with the compact
general expression: the coefficients multiplying \(Q_{K}\), \(C_{2}(G_{K})\,Q_{K}\),
\(C_{2}^{2}(G_{K})\,Q_{K}\), the mixed \(C_{2}(G_{K})\,C(R_{aL})\) piece, and the
\(C(R_{aM})\,C(R_{aN})\) combination coincide. The Yukawa sector also matches in both normalization and
tensor content: the quadratic term \(\sum T_{aK}\,\lambda^{\dagger}\lambda\) and the quartic piece
\(-\tfrac{1}{16\pi^{4}}\sum \lambda^{\dagger}\lambda\,\lambda^{\dagger}\lambda\) take the same form in the
two representations. Crucially, the mixed gauge--Yukawa coefficients appear exactly as \((1+A-B)\) and
\((1+B-A)\) for
\(\sum_{abc,L}\alpha_{0L}\,T_{aK}C(R_{aL})\,\lambda_{0}^{\dagger abc}\lambda_{0\,abc}\) and
\(\sum_{abc,L}\alpha_{0L}\,T_{aK}\,\lambda_{0}^{\dagger abc}C(R_{cL})\lambda_{0\,abc}\), respectively (see
the middle lines of Eq.~\eqref{eq:beta_bare_full}); this is the expected HCD pattern (differences of
regulator constants rather than ratios).

\subsection{Three-Loop RG Evolution with Regulator and Yukawa Contributions}

\begin{table}[ht]
\centering
\begin{tabular}{cc}
\toprule
\textbf{Energy Scale} $\bm{\mu}$ [GeV] & \textbf{Gauge Coupling} $\bm{\alpha_K(\mu)}$ \\
\midrule
$10^3$  & 1.8351 \\
$10^6$  & 1534.8452 \\
$10^9$  & 1534.8452 \\
$10^{12}$ & 1534.8452 \\
$10^{15}$ & 1534.8452 \\
\bottomrule
\end{tabular}
\caption{Sample values of \(\alpha_K(\mu)\) at representative scales corresponding to
Fig.~\ref{fig:RG_Yukawa_Regulator}. Numbers are schematic and reflect the chosen parameters.}
\label{tab:RG_Yukawa_Samples}
\end{table}

The trajectory in Fig.~\ref{fig:RG_Yukawa_Regulator}, together with Table~\ref{tab:RG_Yukawa_Samples},
illustrates how regulator-sensitive gauge and Yukawa contributions can substantially affect the flow
through finite, scheme-dependent terms. These examples underscore the utility of exponential regulators
for analytic control and emphasize that NSVZ compatibility is preserved at the bare level, while finite
redefinitions are needed to expose it in specific renormalization schemes.

% --- Finite Renormalizations and Restoration of NSVZ ---
\section{Finite Renormalizations and Restoration of the NSVZ Structure}
\label{sec:FiniteRenorm}

\begin{figure*}[t]
  \centering
  \includegraphics[width=\linewidth]{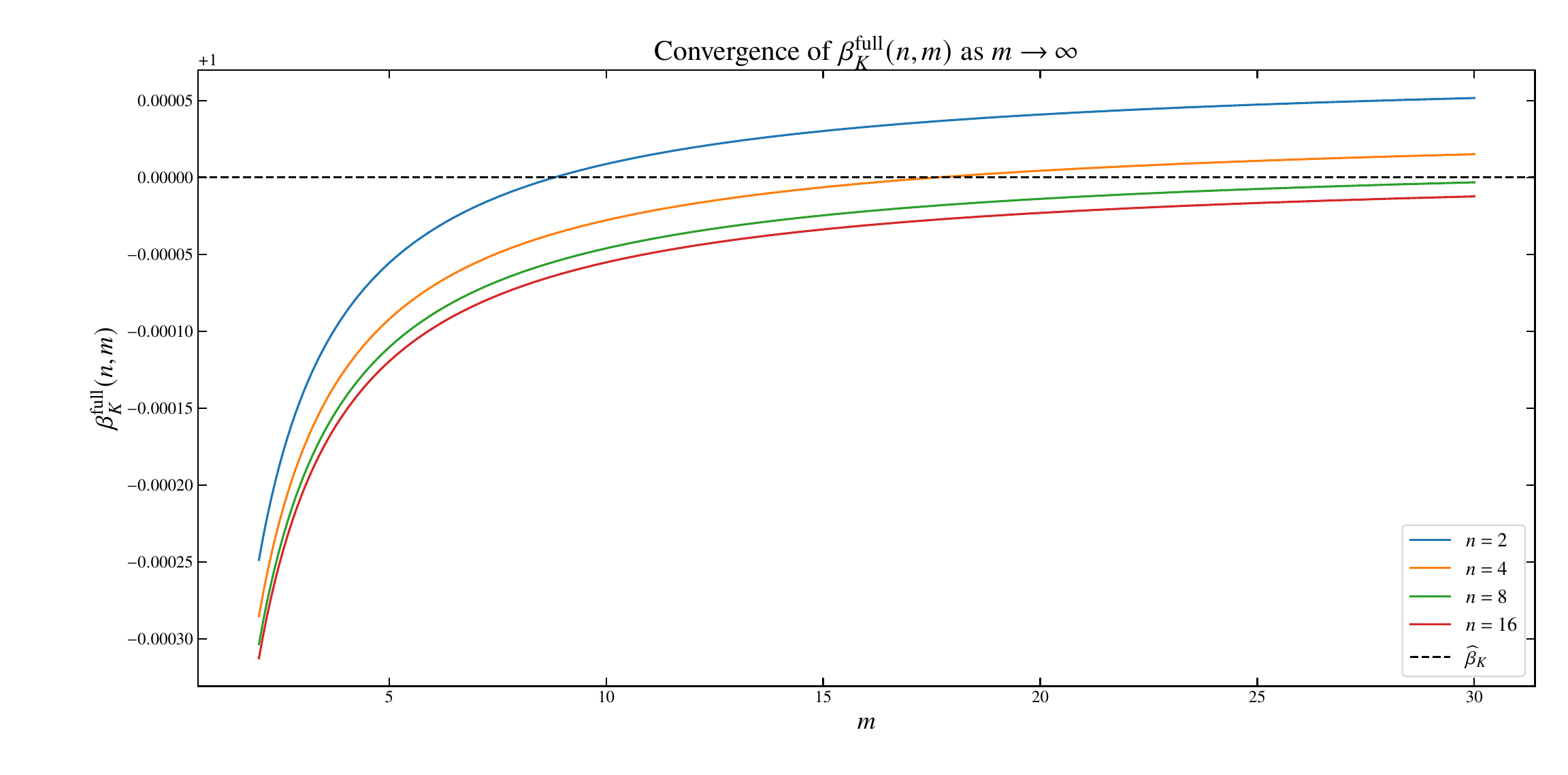}
  \caption{
    Convergence of \(\beta_K^{\text{full}}(n,m)\) to the universal value \(\widehat{\beta}_K\) as
    \(m\to\infty\) for representative fixed \(n\). The leading regulator correction tracks the first
    non-vanishing terms in the \(A,B\) expansion.
  }
  \label{fig:BetaK_convergence_m}
\end{figure*}

\begin{figure*}[t]
  \centering
  \includegraphics[width=\linewidth]{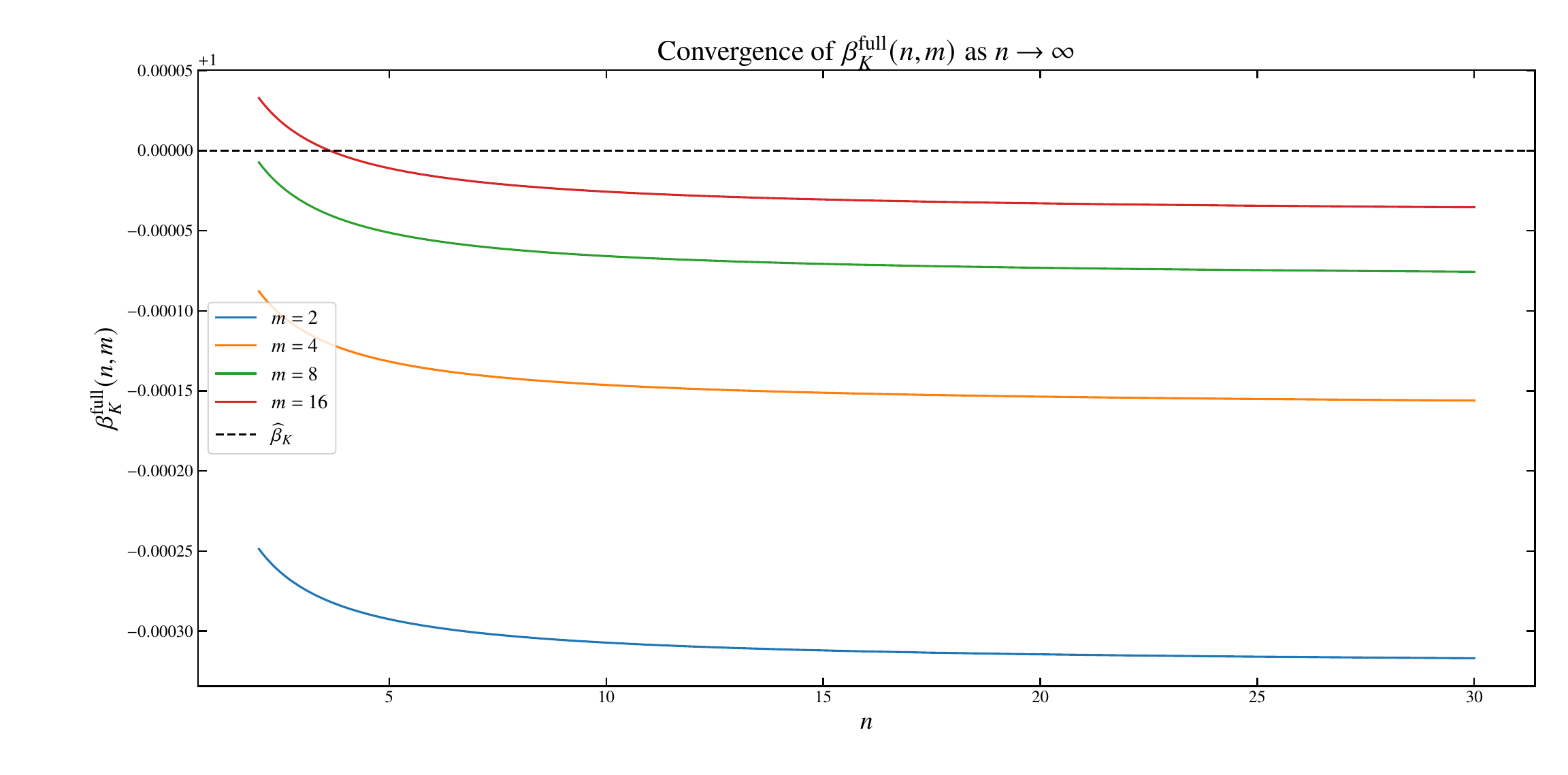}
  \caption{
    Convergence of \(\beta_K^{\text{full}}(n,m)\) to \(\widehat{\beta}_K\) as \(n\to\infty\) for
    representative fixed \(m\). Regulator-dependent terms decay smoothly, leaving the universal flow.
  }
  \label{fig:BetaK_convergence_n}
\end{figure*}

\subsection{Finite Renormalizations of the Couplings}
\label{subsec:finite_redefs}

Bare couplings defined with higher covariant derivative (HCD) regularization satisfy the NSVZ relation
by construction, whereas renormalized couplings in practical schemes such as
\(\overline{\mathrm{DR}}\) need not display the NSVZ form beyond two loops. The difference is entirely due
to \emph{finite}, scheme-dependent contributions that reshuffle higher-loop terms among tensor
structures. One can nevertheless restore the NSVZ form for \emph{renormalized} couplings by applying
finite, analytic redefinitions of the gauge couplings (and, when desired, of Yukawas)
\cite{Siegel1979,Jack1996_1,Jack1996_2,Shifman1996,Kataev2013,Stepanyantz2020}.

Let \(\alpha_K\equiv g_K^2/(4\pi)\) denote the renormalized gauge coupling for the factor \(G_K\).
Introduce a finite, non-singular map
\begin{equation}
\label{eq:finite_map_intro}
  \alpha'_K \;=\; \alpha_K + \delta\alpha_K(\alpha,\lambda),
\end{equation}
with \(\delta\alpha_K\) regular at the origin and expandable as a formal series. Under
Eq.~\eqref{eq:finite_map_intro}, renormalized \(\beta\)-functions transform with the Jacobian
\begin{equation}
\label{eq:Jacobian_beta}
  \widetilde{\beta}'_K(\alpha',\lambda) \;=\;
  \sum_{L}\frac{\partial \alpha'_K}{\partial \alpha_L}\,\widetilde{\beta}_L(\alpha,\lambda)
  \;+\;
  \sum_{a,b,c}\frac{\partial \alpha'_K}{\partial \lambda_{abc}}\,\widetilde{\beta}_{\lambda_{abc}}(\alpha,\lambda).
\end{equation}
Our goal is to choose \(\delta\alpha_K\) such that \(\widetilde{\beta}'_K\) reproduces the NSVZ denominator
structure \(\bigl(1-\tfrac{C_2(G_K)\alpha'_K}{2\pi}\bigr)^{-1}\) through three loops, matching
Eq.~\eqref{eq:NSVZ_three_loop} when written for renormalized couplings.

\paragraph{Parameterization through three loops.}
A convenient parameterization that suffices for three-loop gauge \(\beta\)-functions is
\begin{align}
\label{eq:delta_alpha_param}
  \delta \alpha_K 
  \;=\; & \frac{1}{2\pi}\sum_{L} c^{(1)}_{KL}\,\alpha_K\alpha_L \notag \\
  & \quad + \frac{1}{(2\pi)^2}\,c^{(2)}_{K}\,\alpha_K^2 \notag \\
  & \quad + \frac{1}{(2\pi)^2}\,s^{(1)}_{K\,abc}\,\alpha_K\,\lambda^{abc}\lambda_{abc}^{\dagger} \notag \\
  & \quad + \mathcal{O}(\alpha^4,\alpha^2\lambda^2).
\end{align}
For the \emph{purely gauge} matching displayed below, one may set \(s^{(1)}_{K\,abc}=0\). If one also wishes
to remove finite mixed gauge--Yukawa remnants in a particular scheme, a suitable (tensorial) choice of
\(s^{(1)}_{K\,abc}\) accomplishes that without affecting scheme-invariant combinations.

Matching Eq.~\eqref{eq:Jacobian_beta} to the NSVZ pattern using the \(\overline{\mathrm{DR}}\) result
\eqref{eq:BetaDR} yields a convenient solution for the gauge-only case,
\begin{align}
\label{eq:c1_full}
  c^{(1)}_{KL} &= 
  \begin{cases}
  -\tfrac{1}{4} C_2(G_K), & \text{if } K=L,\\[2pt]
  \phantom{-}\tfrac{1}{4}\displaystyle\sum_a T_{aK}\,C(R_{aL}), & \text{if } K\neq L,
  \end{cases}
\\[4pt]
\label{eq:c2_full}
  c^{(2)}_{K} &= -\tfrac{1}{4}\,Q_K,
\end{align}
with \(Q_K=\sum_a T_{aK}-3\,C_2(G_K)\). These coefficients are fixed by matching to the finite integration
constants \((b_{1,K},b_{2,K},b_{2,KL})\) in Eqs.~\eqref{eq:bDR_definitions}–\eqref{eq:BetaDR}; they are
independent of the particular (admissible) Pauli--Villars (PV) spectrum, which shifts only finite parts
uniformly. Inserting Eqs.~\eqref{eq:c1_full}–\eqref{eq:c2_full} into Eq.~\eqref{eq:delta_alpha_param} gives
\begin{align}
\label{eq:delta_alpha_explicit}
  \delta \alpha_K
  \;=\; \frac{1}{2\pi}\Biggl[ &
     -\frac{1}{4}\,C_2(G_K)\,\alpha_K^2 \notag\\
     &\quad + \sum_{L\neq K}\frac{1}{4}\!\left(\sum_a T_{aK}C(R_{aL})\right)\alpha_K\alpha_L \notag\\
     &\quad - \frac{1}{4}\,Q_K\,\alpha_K^2
  \Biggr] \;+\; \mathcal{O}(\alpha^3,\alpha^2\lambda^2).
\end{align}
With Eq.~\eqref{eq:delta_alpha_explicit}, the renormalized \(\overline{\mathrm{DR}}\) result maps to an
NSVZ-compatible scheme through three loops in the gauge sector. This transformation alters only finite
terms; scheme-invariant combinations at three loops remain unchanged
\cite{Jack1996_1,Jack1996_2,Shifman1996}.

\subsection{Asymptotic Behavior and Regulator-Dependent Terms}
\label{sec:LogResummation}

The HCD regulator functions \(R(x)=e^{x^n}\) and \(F(x)=e^{x^m}\) introduce \emph{finite} regulator
dependence via the constants \(A(n)=\gamma_E/n\) and \(B(m)=(\gamma_E+\ln 2)/m\), see
Eq.~\eqref{eq:A_B_values}. In practical formulae these appear in the combinations \(1+\tfrac{A}{2}\),
\(1+A-B\), and \(1+B-A\), for example in Eq.~\eqref{eq:beta_bare_full}. Since the PV masses are \emph{free}
regularization parameters (subject to the usual consistency conditions), their values affect \emph{only}
finite parts and can be chosen for algebraic convenience without altering scheme-invariant content
\cite{Slavnov1977,Kataev2013}.

It is natural to organize regulator dependence directly in the small parameters \(A(n)\) and \(B(m)\).
Writing the bare \(\beta\)-function as
\begin{equation}
\label{eq:beta_AB_expansion}
  \beta_K(\alpha_0,\lambda_0;n,m)
  \;=\;
  \sum_{r,s\ge 0} A(n)^{r}\,B(m)^{s}\;\mathcal{B}_K^{(r,s)}(\alpha_0,\lambda_0),
\end{equation}
the leading regulator-controlled structures are
\begin{align}
\label{eq:beta_AB_leading}
\mathcal{B}_K^{(1,0)} &\;\propto\; 
   \sum_{a,L}\frac{\alpha_{0L}^2}{2\pi^2}\,T_{aK}C(R_{aL})\,Q_L, \\
\mathcal{B}_K^{(0,1)} &\;\propto\; 
   \sum_{abc,L}\frac{\alpha_{0L}}{8\pi^3}\,T_{aK}C(R_{aL})\,
   \lambda_0^{\dagger abc}\lambda_{0\,abc},
\end{align}
with higher orders generated by further insertions of \(A\) and \(B\). A compact representation is
\begin{align}
  \beta_K^{\mathrm{resum}}
  \;=\; & \sum_{a,L}\frac{\alpha_{0L}^2}{2\pi^2}\,T_{aK}C(R_{aL})\;
  \Phi_{KL}\!\bigl(A(n),B(m)\bigr), \\
  \Phi_{KL}(A,B) 
  \;=\; & Q_L\sum_{r,s\ge 0} \tilde c^{(KL)}_{rs}\,A^{r}B^{s},
\end{align}
where \(\tilde c^{(KL)}_{10}\) and \(\tilde c^{(KL)}_{01}\) reproduce the \(\mathcal{B}_K^{(1,0)}\) and 
\(\mathcal{B}_K^{(0,1)}\) structures in Eq.~\eqref{eq:beta_AB_leading}.

\begin{figure*}[t]
  \centering
  \includegraphics[width=\linewidth]{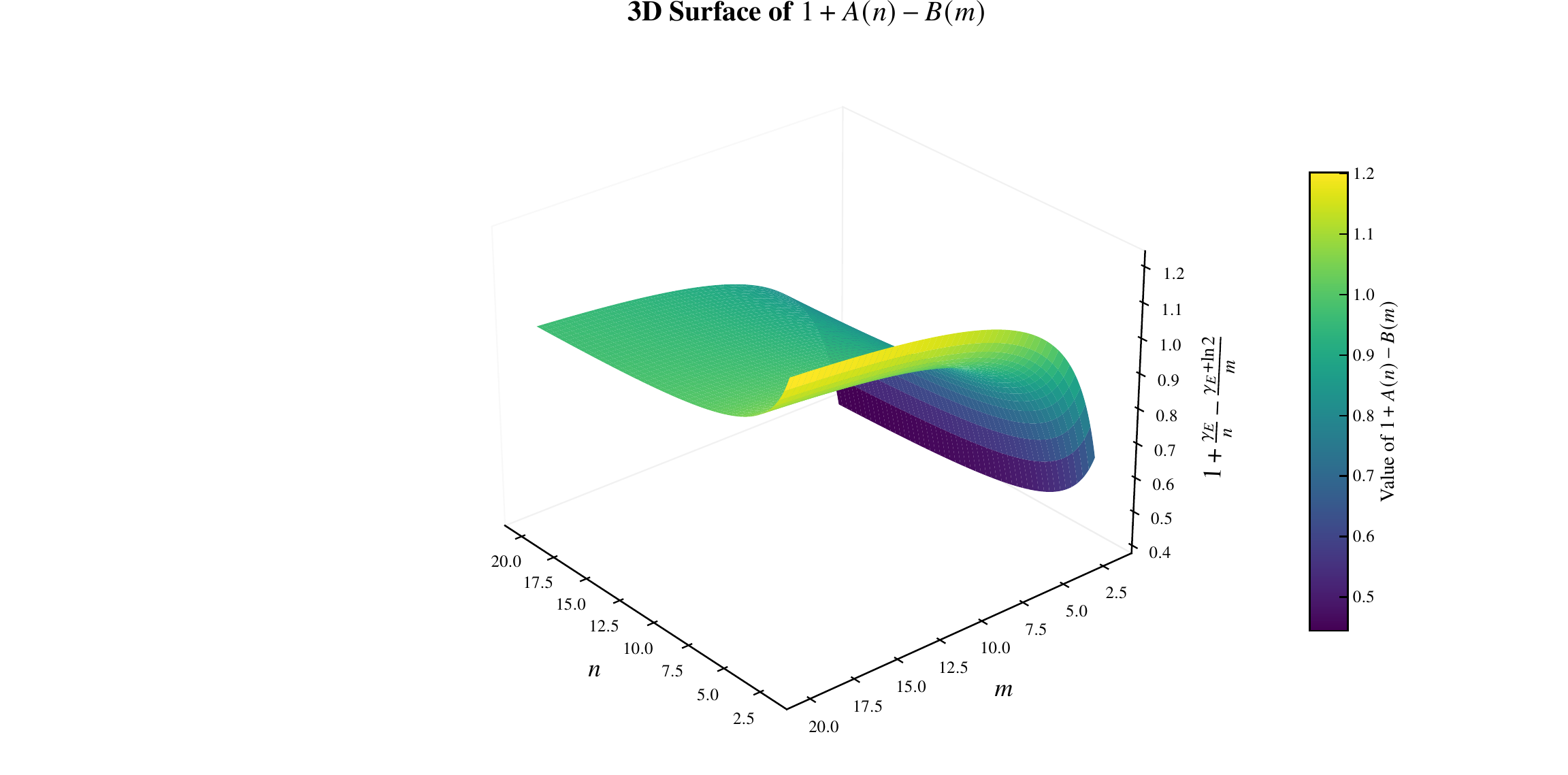}
  \caption{
    Regulator dependence of the mixed gauge--Yukawa coefficient
    \(1+A(n)-B(m)=1+\tfrac{\gamma_E}{n}-\tfrac{\gamma_E+\ln 2}{m}\).
    Increasing \(n\) and \(m\) suppresses the regulator imprint and the coefficient approaches~\(1\),
    as expected when finite, scheme-dependent artifacts are removed.
  }
  \label{fig:RegulatorSurface}
\end{figure*}

Figure~\ref{fig:RegulatorSurface} visualizes the behavior of \(1+A-B\), which multiplies one of the mixed
gauge--Yukawa structures at three loops. The smooth approach to unity as \(n,m\to\infty\) reflects the
suppression of regulator-tagged finite terms.

\subsection{Universal Limit and Scheme Independence}
\label{sec:ScalingLimit}

The full three-loop gauge \(\beta\)-function \(\beta_K^{\text{full}}(n,m)\) approaches a universal,
scheme-independent limit as \(n,m\to\infty\), where \(A(n),B(m)\to 0\) and the regulator-dependent constants
vanish:
\begin{equation}
  \widehat{\beta}_K
  \;=\;
  \lim_{n,m\to\infty}\left[
     \beta_K^{\text{full}}(n,m)
     \;-\;
     \sum_{r+s\ge 1} A(n)^r B(m)^s\,\mathcal{B}_K^{(r,s)}
  \right].
\end{equation}
This limit removes regulator-tagged finite parts and isolates the scheme-invariant content of the
three-loop coefficients; finite scheme changes (including PV choices) do not affect \(\widehat{\beta}_K\)
\cite{Jack1996_1,Jack1996_2,Stepanyantz2020}.

The panels in Figs.~\ref{fig:BetaK_convergence_m} and \ref{fig:BetaK_convergence_n} illustrate how the
finite regulator imprint diminishes as the regulator powers increase. The same mechanism underlies the
\(\overline{\mathrm{DR}}\)\,\(\to\)\,NSVZ scheme transformation: finite redefinitions eliminate regulator-tagged
pieces without affecting scheme-invariant combinations.

\subsection*{Numerical Illustrations of the Yukawa Sector}
\label{subsec:numerics_yukawa}

The next figures visualize the structure and convergence of the Yukawa-enhanced three-loop contribution
\(\beta_K^{(\lambda)}\) in the presence of HCD regulators. The parameter choices are representative and
serve to display typical trends under exponential HCD regulators; conclusions about scheme-invariant
structures do not rely on a specific PV spectrum.

\begin{figure*}[t]
  \centering
  \includegraphics[width=\linewidth]{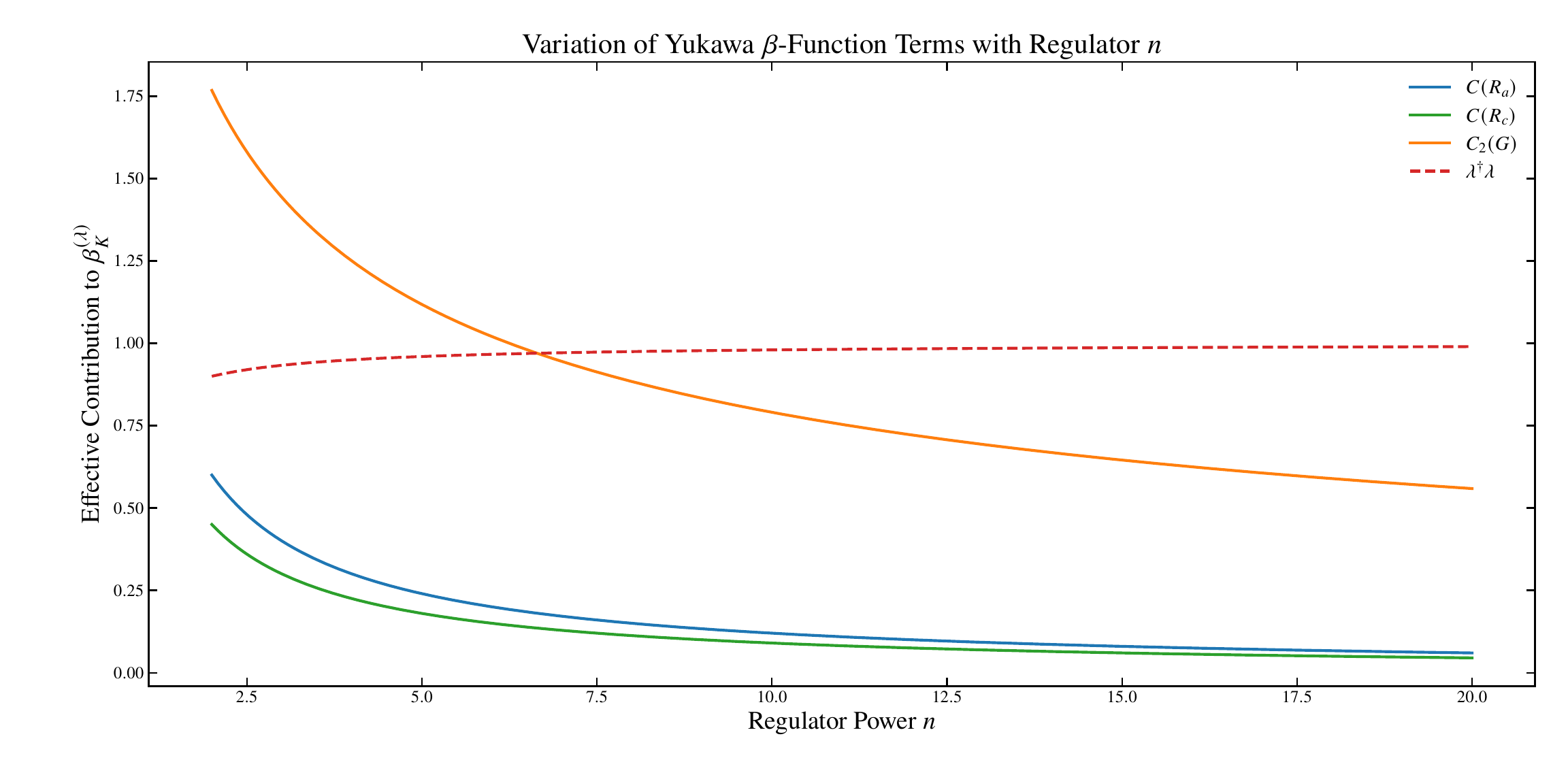}
  \caption{
    Regulator dependence of individual contributions to \(\beta_K^{(\lambda)}\) as a function of the
    gauge-sector power \(n\). Group-theory pieces tied to \(C(R_a), C(R_c), C_2(G)\) decay with increasing
    \(n\), reflecting exponential damping. The self-coupling piece \(\lambda^\dagger\lambda\) is relatively
    stable, indicating its dominance as regulator artifacts diminish.
  }
  \label{fig:YukawaBreakdownLine}
\end{figure*}

\begin{figure*}[t]
  \centering
  \includegraphics[width=\linewidth]{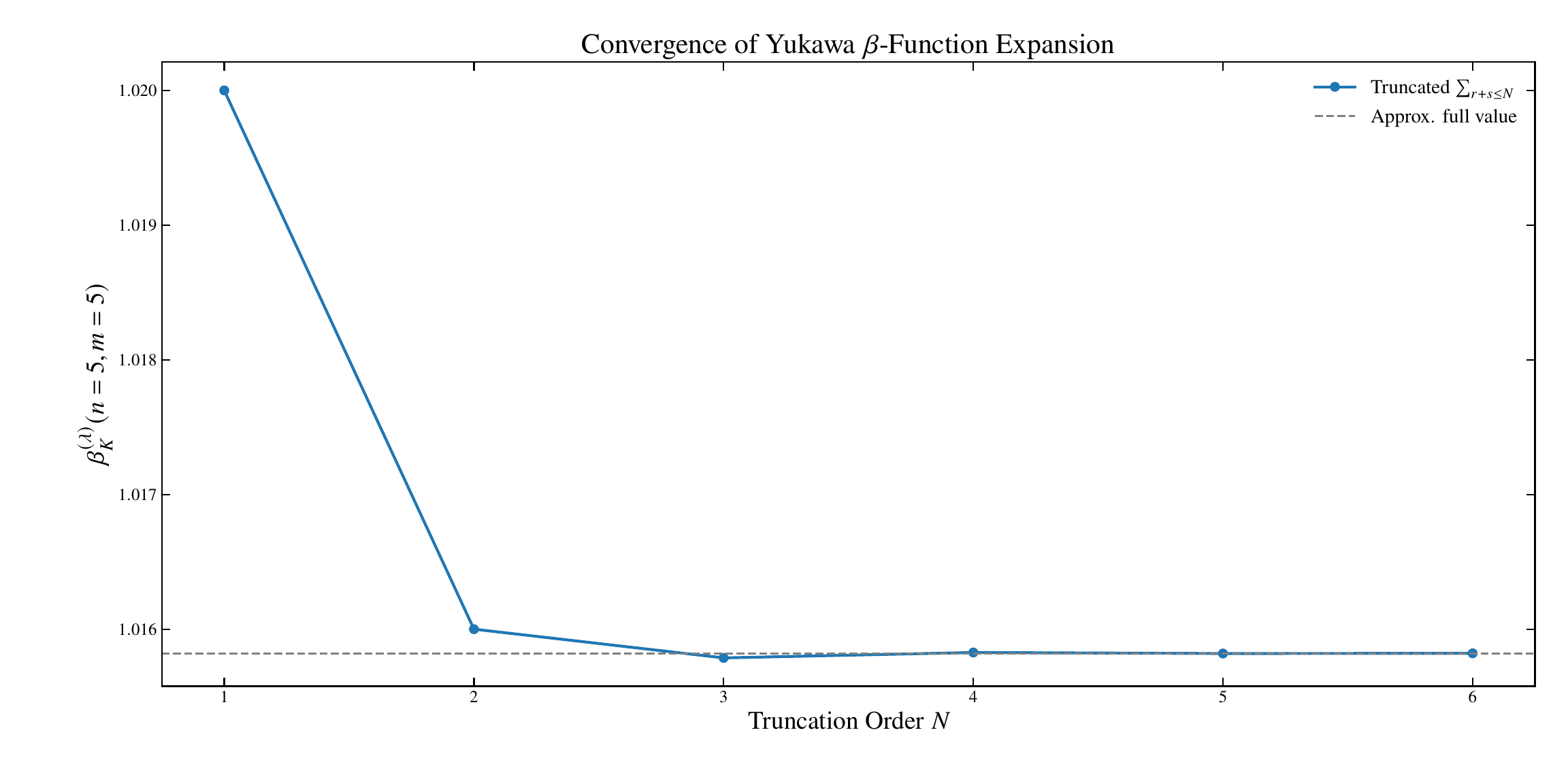}
  \caption{
    Convergence of the truncated expansion
    \(\sum_{r+s\le N} A(n)^r B(m)^s\,\mathcal{B}_K^{(\lambda;r,s)}\) at fixed \((n,m)=(5,5)\).
    The rapid approach to the resummed value shows that low orders capture the essential regulator dependence
    of the Yukawa sector.
  }
  \label{fig:YukawaConvergenceLine}
\end{figure*}

\medskip

In summary, finite redefinitions provide a clean bridge from \(\overline{\mathrm{DR}}\) to an NSVZ-compatible
scheme, while the large-\(n,m\) behavior of exponential HCD regulators makes the regulator imprint on the
three-loop coefficients manifestly controllable. The PV masses function purely as free, auxiliary
parameters: different admissible choices alter only finite parts, leave scheme-invariant combinations
unchanged, and can be employed to simplify intermediate algebra without affecting physical conclusions.

% --- Conclusion ---
\section{Conclusion}
\label{sec:Conclusion}

We have analyzed the three-loop renormalization group (RG) structure of general
$\mathcal{N}=1$ supersymmetric gauge theories within higher covariant derivative (HCD)
regularization supplemented by Pauli--Villars (PV) superfields
\cite{Slavnov1971,Slavnov1972,Slavnov1977,Kataev2013,Stepanyantz2020}.
Working with \emph{bare} couplings---for which the Novikov--Shifman--Vainshtein--Zakharov (NSVZ)
relation is satisfied to all orders in HCD \cite{Novikov1983,Shifman1986,Novikov1986,Stepanyantz2020}---we
identified the finite, regulator-dependent pieces that enter the general three-loop gauge
$\beta$-functions, and computed them in closed form for the exponential regulator family
\(
R(x)=e^{x^{n}},\; F(x)=e^{x^{m}},\; n,m\ge 2
\).
In particular,
\begin{equation}
A(n)=\frac{\gamma_E}{n},\qquad
B(m)=\frac{\gamma_E+\ln 2}{m},\label{eq:ConclusionAB}
\end{equation}
with $\gamma_E$ the Euler--Mascheroni constant \cite{WhittakerWatson}, and we have shown how these
constants multiply the expected group-theory and Yukawa structures in agreement with the general HCD
three-loop formulas \cite{Kazantsev2020,Haneychuk2022,Haneychuk2025}.

Substituting \eqref{eq:ConclusionAB} into the compact three-loop HCD expressions yields fully explicit,
regulator-parameterized bare $\beta$-functions (including Yukawa terms) that manifestly obey the NSVZ form.
We then related these results to renormalized $\overline{\mathrm{DR}}$ couplings
\cite{Siegel1979,Jack1996_1,Jack1996_2}, tracking how the renormalized $\widetilde{\beta}$ differ by \emph{finite}
terms which can be removed by finite coupling redefinitions without affecting any scheme-invariant
combination. In this way, the NSVZ denominator $\bigl(1-C_2(G_K)\alpha_K/(2\pi)\bigr)^{-1}$ is restored for
renormalized couplings through three loops, in line with the holomorphic and anomaly-based logic
underlying NSVZ \cite{Shifman1996,Novikov1983,Novikov1986}.

We compared our exponential-regulator specialization with recent general HCD formulas for multiple gauge
couplings \cite{Haneychuk2025} and found structural agreement after aligning conventions. In particular,
the mixed gauge--Yukawa coefficients appear precisely as $(1+A-B)$ and $(1+B-A)$, which is the characteristic
HCD pattern \cite{Kazantsev2020,Haneychuk2022}. We further clarified the role of PV masses: they are free
regularization parameters constrained only by gauge invariance, supersymmetry, and decoupling; they shift
\emph{only} finite pieces, leaving scheme-invariant three-loop structures intact
\cite{Slavnov1977,Kataev2013}.

Finally, we examined the limit $n,m\to\infty$ where $A(n),B(m)\to 0$ and the regulator-tagged finite pieces
vanish. In this regime the three-loop flow approaches a universal, scheme-independent form, providing both a
useful cross-check and a clean separation between invariant content and scheme artifacts.
Phenomenologically, three-loop corrections reorganize finite matching and inter-gauge mixing effects but do
not drive the GUT scale to the TeV regime; rather, they produce percent-level adjustments that matter for
precision unification and threshold analyses \cite{Ellis1991,Amaldi1991,Langacker1991,Kazakov1998}.

\medskip
\noindent\textit{Outlook.} Natural extensions include: (i) incorporating soft SUSY-breaking and mapping the
resulting scheme transformations in an NSVZ-compatible fashion; (ii) multi-factor gauge theories with large
Yukawa sectors, where finite redefinitions admit a tensorial organization; (iii) applications to IR fixed
points and Seiberg dual pairs, where explicit control of finite parts may sharpen quantitative tests; and
(iv) exploring nonperturbative thresholds and holomorphic contributions in HCD, in light of resurgence
frameworks and trans-series techniques \cite{Aniceto2021,Dunne2016,Marino2008}.

\appendix

% ============================================================
\section{Evaluation of Regulator-Dependent Constants}
\label{app:RegulatorDetails}
% ============================================================

This appendix presents the detailed derivation of the regulator-dependent constants $A(n)$ and $B(m)$
entering the three-loop gauge $\beta$-functions in HCD regularization
\cite{Kataev2013,Kazantsev2020,Haneychuk2022,Stepanyantz2020,Haneychuk2025}.
Although the final results are compact, intermediate expressions contain logarithmically divergent
building blocks whose divergences cancel in the combinations defining $A,B$.

\subsection{Mellin regularization of the master integral}

A recurring object is the logarithmically divergent integral
\begin{equation}
\label{eq:app_master}
\mathcal{I}_p \;\equiv\; \int_0^\infty \frac{dx}{x}\,e^{-x^p}\,,
\end{equation}
which we define by analytic continuation. Introduce a complex regulator $s$,
\begin{equation}
\mathcal{I}_p(s) \;\equiv\; \int_0^\infty x^{s-1} e^{-x^p}\,dx
\;=\; \frac{1}{p}\,\Gamma\!\Bigl(\frac{s}{p}\Bigr)\,,\qquad \Re(s)>0\,.
\end{equation}
Expanding near $s=0$ using
\begin{equation}
\Gamma(\varepsilon) \;=\; \frac{1}{\varepsilon} - \gamma_E
\;+\; \frac{\pi^2}{12}\,\varepsilon + \mathcal{O}(\varepsilon^2)\,,
\end{equation}
we obtain
\begin{equation}
\mathcal{I}_p(s) \;=\; \frac{1}{s} - \frac{\gamma_E}{p}
+ \frac{\pi^2}{12p^2}\,s + \mathcal{O}(s^2)\,,
\end{equation}
so that the finite part is
\begin{equation}
\mathrm{FP}\,\mathcal{I}_p \;=\; \lim_{s\to 0}\!\left(\mathcal{I}_p(s)-\frac{1}{s}\right)
\;=\; -\,\frac{\gamma_E}{p}\,.\label{eq:FP_result}
\end{equation}
This Mellin-regularized prescription makes the finite pieces entering $A,B$ explicit
\cite{WhittakerWatson,Smirnov2004}.

\subsection{Evaluation of \(A(n)\) for \(R(x)=e^{x^n}\)}

The gauge-sector constant is defined by
\begin{equation}
A \;\equiv\; \int_0^\infty dx\,\ln x\,\frac{d}{dx}\!\left(\frac{1}{R(x)}\right)\!,
\qquad R(x)=e^{x^n}\,.
\end{equation}
Thus,
\begin{align}
A(n)
&= \int_0^\infty \ln x \,\frac{d}{dx}\!\Bigl(e^{-x^n}\Bigr)\,dx
= -\,n\int_0^\infty x^{n-1}\ln x\,e^{-x^n}\,dx \notag \\
&\xrightarrow{\,t=x^n\,} -\int_0^\infty \ln t \, e^{-t}\,dt
= \gamma_E
\;\;\Longrightarrow\;\;
\boxed{\,A(n)=\dfrac{\gamma_E}{n}\,}\,,
\end{align}
where integrability at the endpoints ($n\ge 2$) ensures the boundary term vanishes.

\subsection{Evaluation of \(B(m)\) for \(F(x)=e^{x^m}\)}

The matter-sector constant is
\begin{equation}
\begin{split}
B &\equiv \int_0^\infty dx\,\ln x\,
      \frac{d}{dx}\!\left(\frac{1}{F(x)^2}\right),
\\
&\qquad F(x)=e^{x^m}.
\end{split}
\end{equation}

Since
\[
\frac{d}{dx}\!\left(\frac{1}{F(x)^2}\right)
=
\frac{d}{dx}\!\left(e^{-2x^m}\right)
=
-2m\,x^{m-1}e^{-2x^m},
\]
we obtain
\begin{align}
B(m)
&=
-2m \int_0^\infty x^{m-1}\ln x\,e^{-2x^m}\,dx
\nonumber\\
&\xrightarrow{\,u=2x^m\,}
-
\int_0^\infty
\left(
\frac{1}{m}\ln u
-
\frac{\ln 2}{m}
\right)
e^{-u}\,du
\nonumber\\
&=
\frac{\gamma_E+\ln 2}{m}.
\label{eq:B_result_clean}
\end{align}
Each term in the intermediate line is separately divergent; their Mellin-regularized combination is finite
and yields the quoted result. Both $A,B$ agree with the finite constants entering the compact three-loop HCD
formulas \cite{Kazantsev2020,Haneychuk2022,Haneychuk2025}.

\subsection{Scaled exponential profiles}

For scaled profiles $R(x)=e^{c x^p}$, $F(x)=e^{c x^q}$ with $c>0$,
\begin{equation}
\int_0^\infty x^{s-1}e^{-c x^p}dx
= \frac{1}{p}\,\Gamma\!\Bigl(\tfrac{s}{p}\Bigr)\,c^{-s/p}
\;=\; \frac{1}{s} - \frac{\gamma_E+\ln c}{p} + \cdots,
\end{equation}
so that
\begin{equation}
A(p;c)=\frac{\gamma_E+\ln c}{p},\qquad
B(q;c)=\frac{\gamma_E+\ln(2c)}{q}\,.
\end{equation}
These constants differ only by finite, scheme-tagging $\ln c$ shifts, as expected.

\subsection{Auxiliary identity (finite-part form)}

A useful auxiliary identity is
\begin{equation}
\int_0^\infty \frac{dx}{x^{1-s}}\Bigl(1-e^{-x^p}\Bigr)
= \Bigl(\tfrac{1}{s}-\tfrac{1}{s}\Bigr) + \frac{\gamma_E}{p}
- \frac{s}{2p^2} + \mathcal{O}(s^2)\,,
\end{equation}
where the $1/s$ poles cancel explicitly; the finite part is $\gamma_E/p$, consistent with
Eq.~\eqref{eq:FP_result}.

% ============================================================
\section{Pauli--Villars Masses}
\label{app:PVmasses}
% ============================================================

In HCD, Pauli--Villars superfields are introduced to eliminate residual one-loop divergences
\cite{Slavnov1977,Stepanyantz2020}. Their masses are proportional to the UV scale $\Lambda$ and enter only
through the \emph{ratios} $a_{\varphi,K}$ (gauge PV) and $a_{K}$ (matter PV). These ratios are \emph{free}
regularization parameters constrained by gauge invariance, supersymmetry, and decoupling. Different
admissible choices shift only finite parts of multi-loop quantities and do not modify any scheme-invariant
combinations \cite{Kataev2013}. We keep $a_{\varphi,K}$ and $a_K$ symbolic to display finite terms and
scheme dependence transparently.

% ============================================================
\section{Asymptotic Behavior of Yukawa Contributions}
\label{app:YukawaAsymptotics}
% ============================================================

The Yukawa-dependent piece of the three-loop \emph{bare} $\beta$-function admits an expansion in the small
parameters $A(n)=\gamma_E/n$ and $B(m)=(\gamma_E+\ln 2)/m$:
\begin{equation}
\beta_K^{(\lambda)}(\alpha_0,\lambda_0;n,m)
= \sum_{r,s\ge 0} A(n)^{r}\,B(m)^{s}
\,\beta_{K}^{(\lambda;r,s)}(\alpha_0,\lambda_0)\,,
\end{equation}
where the leading nontrivial terms are proportional to $\gamma_E/n$ and $(\gamma_E+\ln 2)/m$ and multiply
the standard gauge--Yukawa tensors (e.g.\ $T_{aK}C(R_{aL})$ and $T_{aK}\,\lambda^\dagger\lambda$),
consistent with \cite{Kazantsev2020,Haneychuk2022}. In the regulator-independent limit
\begin{equation}
\begin{split}
\widehat{\beta}_K^{(\lambda)}
\;\equiv\;
\lim_{n,m\to\infty}
\Bigg[
&\beta_K^{(\lambda)}(\alpha_0,\lambda_0;n,m)
\\
&-
\sum_{r+s\ge 1}
A(n)^r\, B(m)^s\,
\beta_{K}^{(\lambda;r,s)}
\Bigg].
\end{split}
\end{equation}
all finite, regulator-tagged pieces vanish, and one isolates the universal Yukawa contribution. Any residual
finite differences at finite $(n,m)$ can be absorbed by admissible finite redefinitions of couplings
\cite{Jack1996_1,Jack1996_2}, leaving scheme-invariant information unchanged.

\section*{Acknowledgments}

The author thanks Prof. Andrei L. Kataev for useful discussions and valuable correspondence.

\section*{Funding}

This work was funded from the budget of B.M.S. College of Engineering. No additional grants have been received to conduct or direct this particular study.

\section*{Conflict of Interest}

The author declares that he has no conflicts of interest.

\bibliography{RefsGWB}
\bibliographystyle{unsrt}
\end{document}